\documentclass{emulateapj}

\newcommand{\Kepler}{{\it Kepler}}

\newcommand{\thisstar}{KIC 3542116}

\newcommand{\be}{\begin{equation}}
\newcommand{\ee}{\end{equation}}

\newcommand{\kms}{\ensuremath{\rm km\,s^{-1}}}
\newcommand{\ms}{\ensuremath{\rm m\,s^{-1}}}

\usepackage{xcolor}
\usepackage{hyperref}
\usepackage{breakurl}
\usepackage{amsmath}
\usepackage{graphicx}
\usepackage{verbatim}
\usepackage{booktabs}

\usepackage{color}

\slugcomment{}

\shorttitle{Transiting Exocomets}
\shortauthors{Rappaport et al.}

\begin{document}


\title{Likely Transiting Exocomets Detected by {\em Kepler}}
\author {S.~Rappaport\altaffilmark{1}, {A.~Vanderburg\altaffilmark{2,3,4}, T.~Jacobs\altaffilmark{5}, D. ~LaCourse\altaffilmark{6}, J.~Jenkins\altaffilmark{7}, A. Kraus\altaffilmark{8}, A. Rizzuto\altaffilmark{8}, D.\,W.~Latham\altaffilmark {2}, A.~Bieryla\altaffilmark{2}, M.~Lazarevic\altaffilmark{9}, A. ~Schmitt\altaffilmark{10} 
} } 


\altaffiltext{1}{Department of Physics, and Kavli Institute for Astrophysics and Space Research, Massachusetts Institute of Technology, Cambridge, MA 02139, USA, sar@mit.edu}

\altaffiltext{2}{Harvard-Smithsonian Center for Astrophysics, 60 Garden Street, Cambridge, MA 02138 USA; avanderburg@cfa.harvard.edu} 
\altaffiltext{3}{Department of Astronomy, The University of Texas at Austin, 2515 Speedway, Stop C1400, Austin, TX 78712} 
\altaffiltext{4}{NASA Sagan Fellow} 

\altaffiltext{5}{12812 SE 69th Place Bellevue, WA 98006; tomjacobs128@gmail.com, USA}

\altaffiltext{6}{7507 52nd Place NE Marysville, WA 98270; daryll.lacourse@gmail.com, USA}

\altaffiltext{7}{NASA Ames Research Center, Moffett Field, CA 94035, USA}

\altaffiltext{8}{Department of Astronomy, University of Texas, Austin, 78712-1205, USA}

\altaffiltext{9}{Department of Physics, Northeastern University, 100 Forsyth St, Boston, MA 02115}

\altaffiltext{10}{Citizen Scientist, aschmitt@comcast.net , USA}

{}


\begin{abstract}
We present the first good evidence for exocomet transits of a host star in continuum light in data from the \Kepler\ mission. The {\em Kepler} star in question, KIC 3542116, is of spectral type F2V and is quite bright at $K_p = 10$. The transits have a distinct asymmetric shape with a steeper ingress and slower egress that can be ascribed to objects with a trailing dust tail passing over the stellar disk.  There are three deeper transits with depths of $\simeq 0.1\%$ that last for about a day, and three that are several times more shallow and of shorter duration.  The transits were found via an exhaustive visual search of the entire {\em Kepler} photometric data set, which we describe in some detail. We review the methods we use to validate the {\em Kepler} data showing the comet transits, and rule out instrumental artefacts as sources of the signals.  We fit the transits with a simple dust-tail model, and find that a transverse comet speed of $\sim$35-50 km s$^{-1}$ and a minimum amount of dust present in the tail of $\sim 10^{16}$ g are required to explain the larger transits.  For a dust replenishment time of $\sim$10 days, and a comet lifetime of only $\sim$300 days, this implies a total cometary mass of $\gtrsim 3 \times 10^{17}$ g, or about the mass of Halley's comet. We also discuss the number of comets and orbital geometry that would be necessary to explain the six transits detected over the four years of {\em Kepler} prime-field observations. Finally, we also report the discovery of a single comet-shaped transit in KIC 11084727 with very similar transit and host-star properties.

\end{abstract}


\section{Introduction}
\label{sec:intro}

Advances in both space-based missions and ground-based observational techniques over the past dozen years have led to a huge expansion in the number of confirmed exoplanet detections. Currently, there are over 3500 exoplanets confirmed to orbit a variety of host star spectral types. Growing catalogs of short-period transiting exoplanets derived from data returned by the CoRoT (Baglin et al.~2006) and {\em Kepler} (Borucki et al.~2010) spacecraft are complementing a census of longer period objects being compiled from radial velocity and microlensing campaigns  (Mayor \& Queloz 1995; Marcy et al.~1997; Bond et al.~2004). Despite these successes, relatively little is known about the populations of extrasolar minor bodies within these systems (e.g., planetesimals, asteroids and comets). While planet formation theories generally predict that such minor bodies are a ubiquitous byproduct of protoplanetary disk evolution and should be found on scales loosely analogous to those observed in the solar system, their low masses and small radii present extreme challenges to detection via solid-body transits and radial velocity techniques. Even in the most favourable cases, the detection of extrasolar minor bodies in either radial velocity variations or solid-body transits would require sensitivity orders of magnitude higher than the current state of the art.

Presently, the smallest solid-body which has been detected in transit is Kepler-37b, a 0.27\,$R_\oplus$ object on a 13-day period around a solar-like main sequence star (Barclay et al.~2013). The smallest object detected via its host star's reflex motion is the lunar-mass PSR B1257+12 d, detected via exquisitely sensitive pulsar-timing observations \citep{wolszczan}. In some cases, it is possible to detect even smaller sized objects in white-light transit observations because these objects are surrounded by optically thick material (e.g., dust) which increases the transit depths. Examples of such smaller objects include the so-called ``disintegrating planets'' (KIC 12557548b, aka `KIC 1255b' or KOI 3794, Rappaport et al.~2012; KOI 2700b, Rappaport et al.~2014a; K2-22b, Sanchis-Ojeda et al.~2015), which have been detected in transit. It is believed that these are rocky bodies of lunar size or smaller (Perez-Becker \& Chiang 2013) in short-period orbits (9-22 hrs) that produce transits only by virtue of the dusty effluents that they emit (van Lieshout \& Rappaport 2017). A perhaps similar scenario has been detected for the white dwarf WD 1145+017 (Vanderburg et al.~2015). This is an isolated white dwarf that is being orbited by debris with periods of $\sim$4.5-5 hours, which apparently emit dusty effluents that can block up to 60\% of the star's light (G\"ansicke et al.~2016; Rappaport et al.~2016; Gary et al.~2017). It is currently unknown how small the involved bodies are, but estimates range from the mass of Ceres on down. 

There are other avenues to studying extrasolar minor planets that lie outside of the traditional exoplanet detection methods. Radio observations, in particular detections of circumstellar CO emission around stars such as HD 181327 (Marino et al.~2016), Eta Corvi (Marino et al.~2017), and Fomalhaut (Matr\`a et al.~2017) have been attributed to the presence of substantial populations of minor bodies at large orbital separations. Another sensitive method for detecting and understanding populations of extrasolar minor planets is through time-series spectroscopy, rather than time-series photometry (e.g., as performed by \Kepler). Growing evidence for the existence of large populations of extrasolar minor bodies orbiting other stars has come from the detection of anomalous absorption features in the spectra of at least 16 A-type stars where Falling Evaporating Bodies (`FEBs') are proposed to randomly cross the observing line of sight (e.g., Ferlet et al.~1987; Beust et al.~1990; Welsh \& Montgomery 2015). FEBs can be classified as planetesimals or exocomets which have been perturbed into eccentric orbits resulting in a star-grazing periapsis and significant sublimation of volatiles within $\lesssim 0.5$ AU of the stellar photosphere.  This phenomenon results in variable, often red-shifted, absorption-line features typically superposed on the CaII H \& K photospheric lines. Such features have been demonstrated to manifest themselves on short time scales (hours to days) for a number of known FEB systems associated with young A-B type stars including Beta Pictoris (Smith \& Terrile 1984), 49 Ceti (Zuckerman \& Song 2012), HD 42111 (Welsh \& Montgomery 2013), HD 172555 (Kiefer et al.~2014a) and Phi Leonis (Eiroa et al.~2016). Beta Pictoris itself represents an important benchmark system as it is young ($\sim$23 Myr), hosts a massive directly imaged exoplanet, and has also been the target of an extensive 8-year long spectroscopic survey. The latter has revealed that the transient absorption features are bimodal in depth and may arise from two distinct populations of exocomets (Kiefer et al.~2014b). 

In this work we describe the first good evidence for exocomet transits of a host star in continuum light. The object in question is KIC 3542116, a young, magnitude 10, spectral type F2V star observed from 2009 to 2013 by the {\em Kepler} mission. The paper is organised as follows: In Section \ref{sec:search} we define the search methods and the analysis tools used in the identification of exocomet host-star candidates. In Section \ref{sec:discovery} we present what appear to be comet transits of KIC 3542116 with their distinctly asymmetric profiles. Section \ref{sec:validation} discusses our data validation methods and quality assessment of the archival photometry. Section \ref{sec:groundbased} describes the supplemental information we have gathered regarding the host star KIC 3542116 including its photometric properties, UKIRT image, spectrum, Keck high-resolution imaging, and a study of the 100-year photometric history based on the Harvard Plate Stack collection. Section \ref{sec:model} describes the model fits for the six significant transit events. In Section \ref{sec:interpret} we interpret the model results under a variety of {\em a priori} assumptions and list several possible scenarios to explain the observations. In Section \ref{sec:KIC11084727} we present evidence for a single similar comet-shaped transit in KIC 11084727, a near twin to KIC 3542116. A number of discussion items are presented in Section \ref{sec:discuss}. Section \ref{sec:summary} offers a summary of our work and draws some overall conclusions.

\section{Visual Search of the Kepler Data Set}
\label{sec:search}
    
Much of the {\em Kepler} data base has been thoroughly and exhaustively searched for periodically occurring exoplanet transits (e.g., Borucki et al.~2010; Batalha et al.~2013) and binary eclipses (Pr\v sa et al.~2011; Slawson et al.~2011; Matijevi\v c et al.~2012) yielding some 3000 viable planet candidates and a comparable number of eclipsing binaries. The types of algorithms employed include the \Kepler\ team's Transiting Planet Search \citep[TPS,][]{jenkinstps}, Box Least Squares technique (BLS, Kov\'acs et al.~2002) and Fast Fourier Transforms (`FFTs'; see, e.g., Sanchis-Ojeda et al.~2014).  There have also been a number of searches carried out for single (i.e., `orphaned') exoplanet transits (Wang et al.~2015; Foreman-Mackey et al.~2016; Uehara et al.~2016, Schmitt et al.~2017). Additionally, searches for astrophysical transit signals that are only quasi-periodic have been carried out in an automated way (see, e.g., Carter \& Agol 2013). 
    
\begin{figure*}
\centering
\includegraphics[width=5.5in]{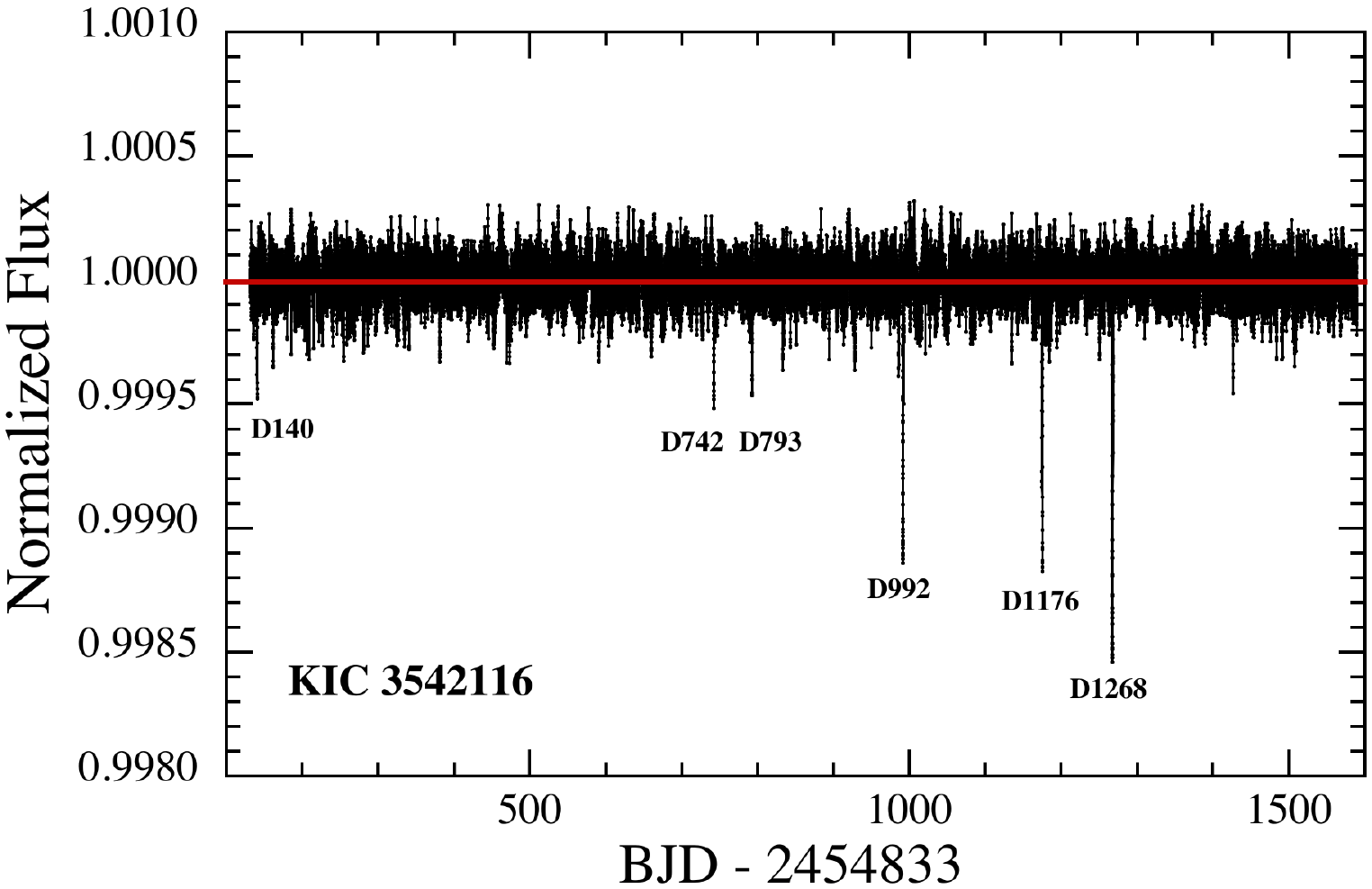} 
\caption{{\em Kepler} PDCSAP photometric lightcurve of KIC 3542116 spanning four years of the main {\em Kepler} mission.  The data train has been harmonically filtered in preparation for searching for planets in the Transiting Planet Search module of the \Kepler\ science pipeline (Jenkins et al.~2010).  The three deepest transits are marked as ``D992'', ``D1176'', and ``D1268'', and have depths ranging from 0.12\% to 0.15\%.  The three more shallow and narrow transits (``D140'', ``D742'', and ``D793'') are partially obscured by some residual $\sim$20-day spot modulations that leak into the photometric aperture from KIC 3542117, and the $\sim$1-day spot rotation period of KIC 3542116.}
   \label{fig:lc}
\end{figure*}  

In an effort to further explore the larger {\em Kepler} data set for isolated transits or aperiodic phenomena, one of us (TJ) undertook a detailed {\em visual} search of the complete Q1-Q17 {\em Kepler} lightcurve archive spanning 201,250 target stars for Data Release 25 (Thompson et al.~2016a) produced by the final Kepler Science Operations Center 9.3 pipeline \citep{jenkinsKDPH2017}. The survey was conducted using the \textsc{LcTools}\footnote{https://sites.google.com/a/lctools.net/lctools/} software system (Kipping et al.~2015), a publicly available Windows-based set of applications designed for processing lightcurves in a fast and efficient manner. Two primary components from the system were utilised; \textsc{LcGenerator} for building lightcurve files in bulk and \textsc{LcViewer} for visually inspecting plots of the lightcurve files for signals of interest.
    
In this survey, lightcurve files were built by \textsc{LcGenerator} in batches of 10,000 files. To build a lightcurve file for a given star, \textsc{LcGenerator} (1) downloaded all available long-cadence time series files from MAST\footnote{https://archive.stsci.edu/kepler/} for Quarters 1-17\footnote{The {\em Kepler} data are downlinked and processed in approximately 90-day ``quarters''.}, (2) extracted the time stamps and \mbox{\textsc{PDCSAP}} flux values from the files excluding data points having a non-zero \textsc{SAP$_-$QUALITY} value, (3) normalised the flux values to a mean value of 1.0, and (4) wrote the combined results to a text file.  
    
To expedite the survey, \textsc{LcViewer} was run concurrently with \textsc{LcGenerator}. As one batch was building, another batch was being inspected. To inspect a batch in \textsc{LcViewer}, a `Work Group' set of text files was first produced. Once a Work Group was established, each file from that set could be opened and displayed sequentially at the click of a button---the process was nearly instantaneous. If the host star had associated Kepler Objects of Interest (KOIs), as obtained from the NASA Exoplanet Archive\footnote{https://exoplanetarchive.ipac.caltech.edu/} (Akeson et al.~2013), the KOI signals were automatically displayed in colour (highlighted) on the viewing screen with annotations when hovering the cursor over a transit signal for easy identification. Signals were highlighted for confirmed planets, planet candidates, and false positives. Any remaining signals were then examined in more detail as possible signals of interest. \textsc{LcViewer} allowed for rapid scrolling through each lightcurve presentation of the entire 17 {\em Kepler} quarters with excellent temporal and flux resolution. 
    
The visual survey of the 201,250 unique {\em Kepler} target lightcurves was conducted over the course of 5 months, beginning in January 2017. Approximately 2000 lightcurves were studied during each active day of the survey, requiring some 5 hours of visual study, allotting about 10 seconds to each target star that showed nothing interesting or unusual. Much more time was spent on the small percentage of stars that revealed one or more potentially interesting features. If an interesting photometric feature was noticed, that object was flagged for further study, vetting, and discussion. 
    
During the course of this comprehensive review, KIC 3542116 was identified as a target of interest due to three anomalous, asymmetric transit-like features occurring in Quarters 10, 12 and 13. These transits were not difficult to spot, with $\gtrsim$ 0.1\% depths and $\sim$1-day durations. The \Kepler\ lightcurve of \thisstar\ and the transits are described in detail in Section \ref{sec:discovery}. We also detected a similar-looking single asymmetric transit in the lightcurve of another target: KIC 11084727. We discuss this object in more detail in Section \ref{sec:KIC11084727}. 
    
In addition to these two stars showing asymmetric transits in their lightcurve, we also identified other objects of interest such as single exoplanet transits and mutual lensing events in binaries. These objects will be discussed in detail in a future paper.

\begin{deluxetable}{lcc}[h!]
\tablewidth{0pt}
\tablehead{
\colhead{Parameter} & \colhead{KIC 3542116} & \colhead{KIC 11084727}}
\startdata
RA (J2000) &  19:22:52.94  & 19:28:41.19 \\  
Dec (J2000) &  38:41:41.5 &  48:41:15.1 \\  
$K_p^a$ & 9.98  & 9.99 \\
$B^b$ & 10.49 & 10.45 \\  
$g^a$ & 10.38  & 10.13 \\
$V^b$ & 10.03 & 10.04 \\  
$r^a$ & 9.99  & 9.94 \\  
$i^a$ & 9.53 & 9.95 \\
$z^a$ & ...  & 10.00 \\
$J^c$ & 9.25 & 9.23 \\
$H^c$ & 9.10 & 9.07 \\
$K^c$ & 9.07 & 9.06 \\
W1$^d$ & 9.06 & 8.98 \\
W2$^d$ & 9.06 & 9.00 \\
W3$^d$ & 8.97 & 9.03 \\
W4$^d$ & 8.30 & 8.49 \\
$T_{\rm eff}^e$ (K) & $6918 \pm 122$ & $6790 \pm 120$ \\ 
$\log \, g^b$ (cgs) & $4.22 \pm 0.12$ & $4.18 \pm 0.19$\\
$M^f$ ($M_\odot$) & $1.47 \pm 0.10$ & $1.45 \pm 0.12$ \\
$R^f$ ($R_\odot$) & $1.56 \pm 0.15$ & $1.55 \pm 0.15$ \\
$[m/H]^e$ & $0.04 \pm 0.11$ & $-0.06 \pm 0.11$ \\
RV$^e$ (km s$^{-1}$) & $-21.1 \pm 0.7$ & $+1.5 \pm 0.5$ \\
Distance$^g$ (pc) & $260^{+30}_{-15}$  & $250^{+30}_{-15}$ \\  
Distance$^h$ (pc) & $235-335$ & $225-255$ \\ 
$v \, \sin i^e$ (km s$^{-1}$) & $57.3 \pm 0.3$ & $32 \pm 0.9$ \\
$\mu_\alpha^{h,b}$ (mas ~${\rm yr}^{-1}$) & $+7.6 \pm 1.1$  &  $+2.5$ \\ 
$\mu_\delta^{h,b}$ (mas ~${\rm yr}^{-1}$) &  $-3.1 \pm 1.1$ & $-22.9$ \\ 
\enddata
\tablecomments{(a) MAST; {\url{http://archive.stsci.edu/k2/data\_search/search.php}}. (b) VizieR \url{http://vizier.u-strasbg.fr/}; UCAC4 (Zacharias et al.~2013). (c) 2MASS catalog (Skrutskie et al.~2006). (d) WISE catalog (Wright et al.~2010; Cutri et al.~2013). (e) TRES spectrum; see Sect.~\ref{sec:TRES}. (f)  Yonsei--Yale tracks (Yi et al.~2001).  (g) Based on photometric parallax only. (h) The Gaia Mission; Prusti et al.~(2016).} 
\label{tbl:mags}
\end{deluxetable}

\section{Discovery of Exocomet Transits in KIC 3542116}
\label{sec:discovery}
        
After \thisstar\ was initially identified as an object of interest, we performed a more thorough inspection of the four-year \Kepler\  lightcurve. \thisstar\ was observed during the entire prime \Kepler\ mission with high photometric precision of about 35 ppm per 30 minute exposure thanks to its bright {\em Kepler}-band magnitude of $K_p = 10$. The full {\em Kepler} lightcurve is shown in Figure \ref{fig:lc}. 

Initially our interest was drawn to the three transit events described in Section \ref{sec:search}. These events are high-signal-to-noise, with depths about 20 times greater than the typical scatter of the \Kepler\ data points. Upon closer inspection, we identified three additional shallower transits with depths about half that of the three deep transits we initially identified. These shallower transits have similar asymmetric profiles to the deep ones, but shorter durations. We label these six dips in the full lightcurve plot shown in Figure \ref{fig:lc}. We label the dips by the date on which they took place (in the \Kepler\ Julian Date reference system BJD - 2454833). The deep dips are labeled D992, D1176, and D1268 and the shallow dips are labeled D140, D742, and D793. 

\begin{figure}
   \centering
   \includegraphics[width=0.99 \columnwidth]{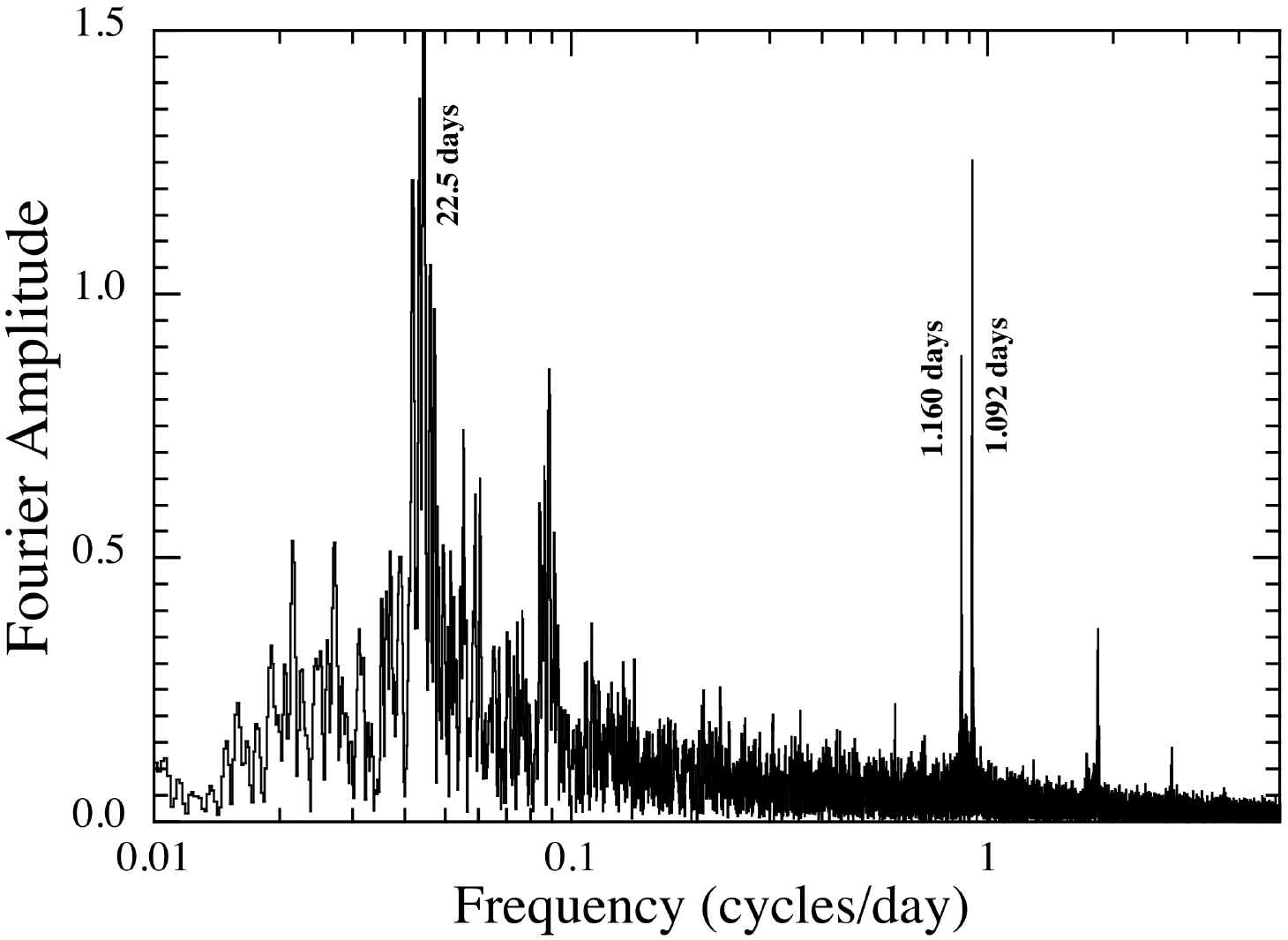} 
   \caption{Fast Fourier transform of the KIC 3542116 PDCSAP photometric time series.  The peaks near 1 day periods, and their harmonics, are due to starspot rotation in KIC 3542116, while the messy peak structure near 23-day periods is leakage from the corresponding spot rotation period in the neighbouring star KIC 3542117 (see also McQuillan et al.~2014).}
   \label{fig:FFT}
\end{figure}

In order to assess the harmonic content in the flux times series of KIC 3542116, we take the Fourier transform of the PDCSAP time series (similar to that shown in Fig.~\ref{fig:lc}). The FFT in Fig.~\ref{fig:FFT} shows two close periods at 1.092 d and 1.160 d, which are likely due to the same underlying rotation period of KIC 3542116; the two periods are most probably due to differential rotation of spots at different stellar latitudes (see, e.g., Reinhold et al.~2013). The $\simeq$ 1 day signals have a semi-amplitude of about 175 ppm. The array of periodicities near 23 days is due to photometric leakage of a spot rotation period in KIC 3542117 (see also McQuillan et al.~2014), a neighbouring star some $10''$ to the north. The 22-23 day signal has a similar semi-amplitude of about 150 ppm. 
                
In order to obtain a clearer view of the six transit events, we attempted to separate the transits from the two rotational signals present in the lightcurve. We were easily able to filter the 22 day period signal by fitting a basis spline to the lightcurve while iteratively excluding outliers and dividing away the best-fit spline. For a more detailed description and illustration of this process, see Figure 3 from \citet{vj14}. It proved more difficult to filter the second rotational signal from the lightcurve because it has a period of about a day, which is similar to the duration of the larger transits we detect around \thisstar. This coincidence of timescales makes it particularly tricky to filter or remove the stellar variability while preserving the transits and not modifying their shapes. We attempted to separate the stellar variability from the transits using both Fourier filtering methods and fitting and removing splines to the data and filtering, but found the results unsatisfactory. We achieved better results filtering the data using Gaussian Process (GP) regression \citep{rasmussen}. In brief, Gaussian process regression involves modelling the covariance properties of a dataset. The learned covariance properties can then be used to predict (either interpolating or extrapolating) how a dataset might behave in the absence of data.  

We took a snippet of the lightcurve around each transit with a duration of about 15 to 20 days and removed both $3-\sigma$ outliers from the lightcurve and data points taken during and around transit. We trained a Gaussian process with a quasi-periodic kernel function \citep[see Equation 4 from][]{haywood} on the lightcurve, optimising the parameters describing the kernel function to best match the lightcurve's covariance properties. We then used our optimised kernel function to predict the behaviour of the stellar activity during the transits, and divided the \Kepler\ lightcurve by the GP prediction to obtain a filtered lightcurve.\footnote{We note that in many cases, for example when dealing with stellar variability in the presence of periodic transits, it is preferable to fit a model to the signal along with a Gaussian process to absorb the stellar variability \citep[see, for example,][]{grunblatt1, grunblatt2}. In our case, however, since we do not {\em a priori} know which models appropriately describe the transits around \thisstar, it is best to attempt to separate the stellar variability from the transits without making assumptions about the shape of the transits.}

\begin{figure*} 
   \centering
   \includegraphics[width=2.3in]{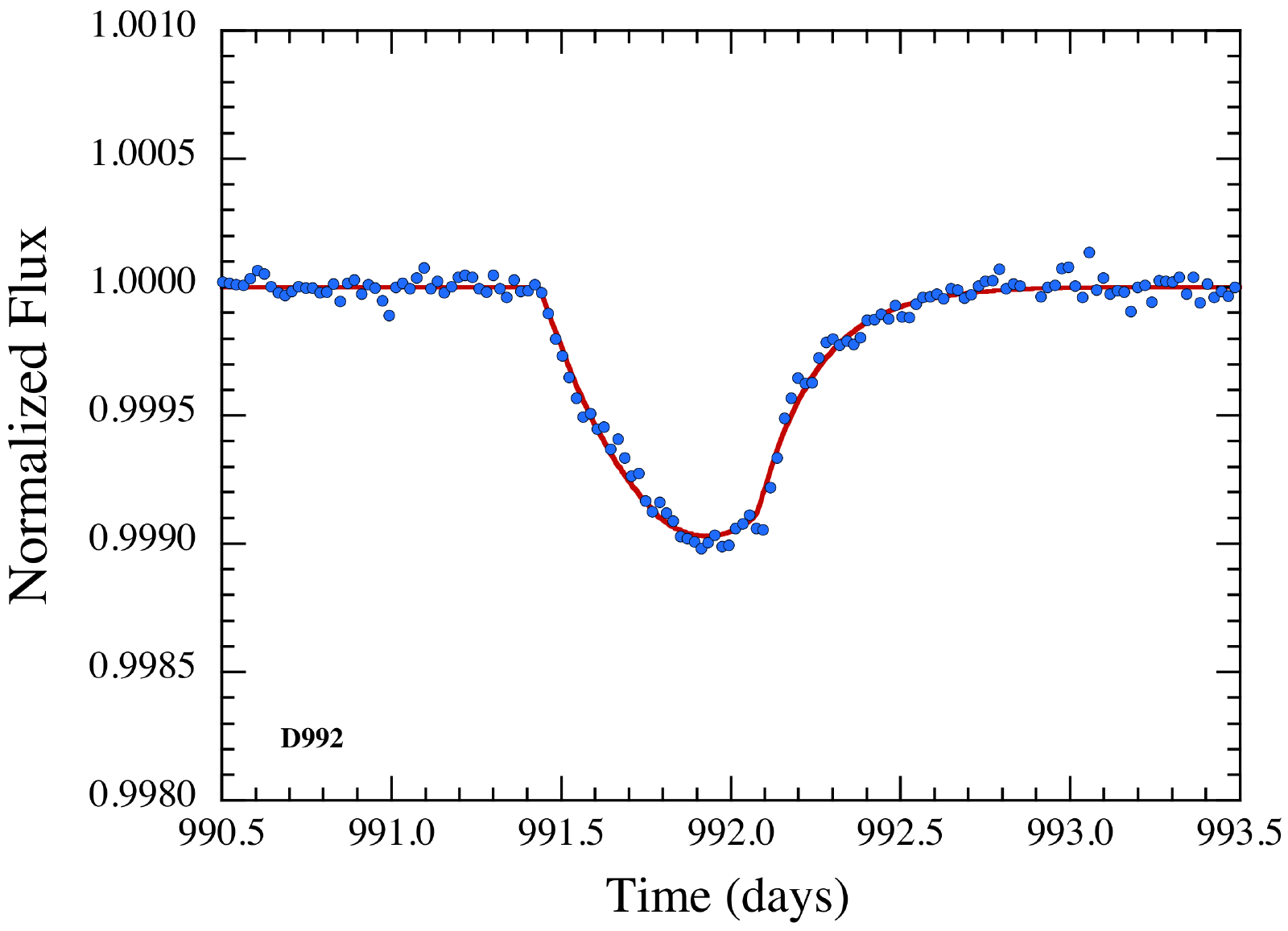} \hglue0.001cm
   \includegraphics[width=2.0in]{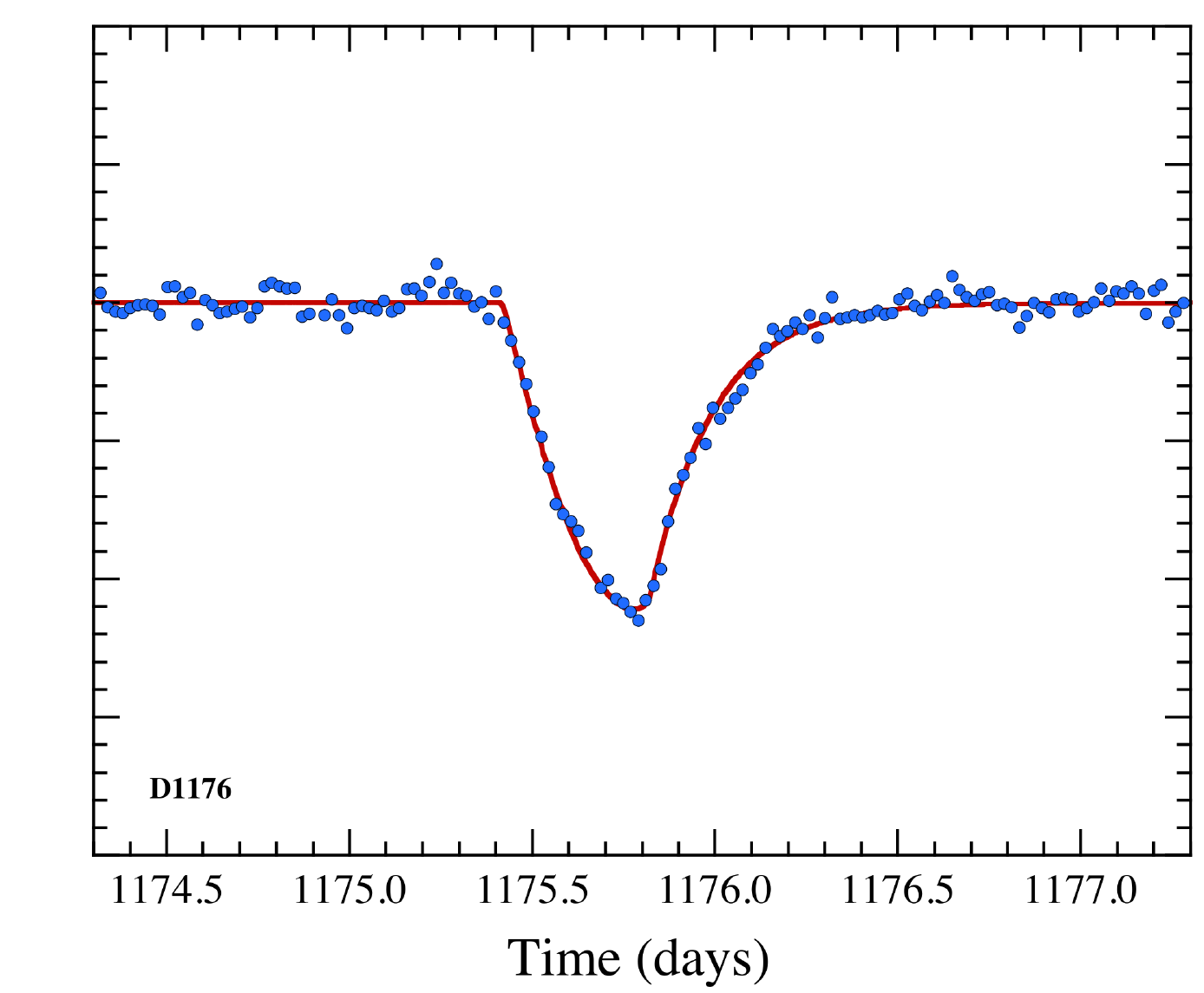} \hglue0.1cm
   \includegraphics[width=2.0in]{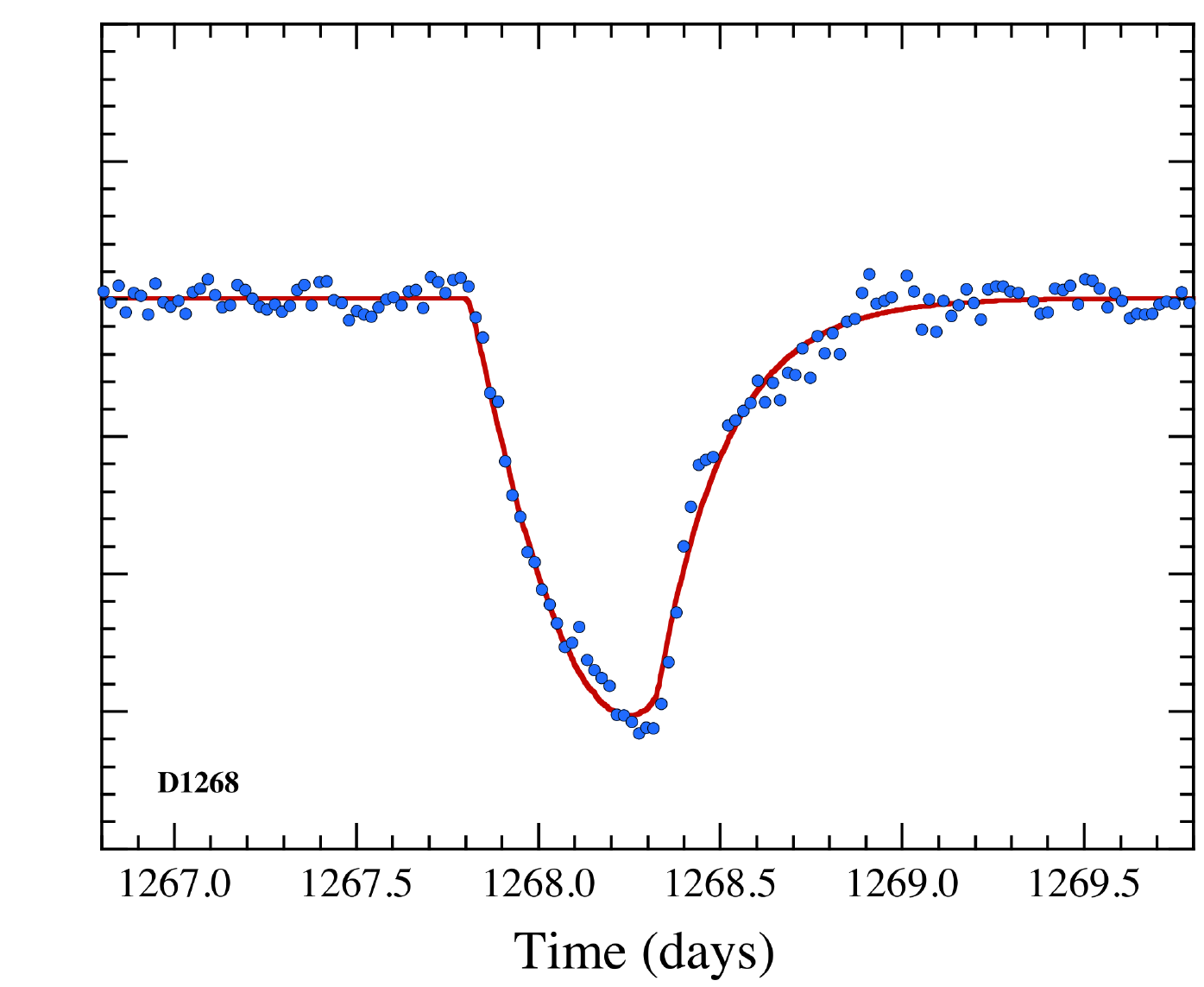} \vglue0.1cm
   \caption{{\em Kepler} SAP photometry covering 3 days around each of the three larger comet transits.  The data have been cleaned via a Gaussian processes algorithm so as to remove most of the 20-day and 1-day spot modulations, as well as other red noise (see text).  The red curves are model fits which will be discussed in Section 5.}
   \label{fig:transit_profiles}
\end{figure*}  

We show the flattened \Kepler\ lightcurve of \thisstar\ around the three deep transits in Figure \ref{fig:transit_profiles}. All three transit profiles have remarkably similar widths, shapes, and depths. In particular, all the transits have steeper ingresses with positive curvature, followed by longer egresses with negative curvature.  The transits are typically 0.12-0.15\% deep and last for about a day. 

\begin{figure*} 
   \centering
   \includegraphics[width=2.3in]{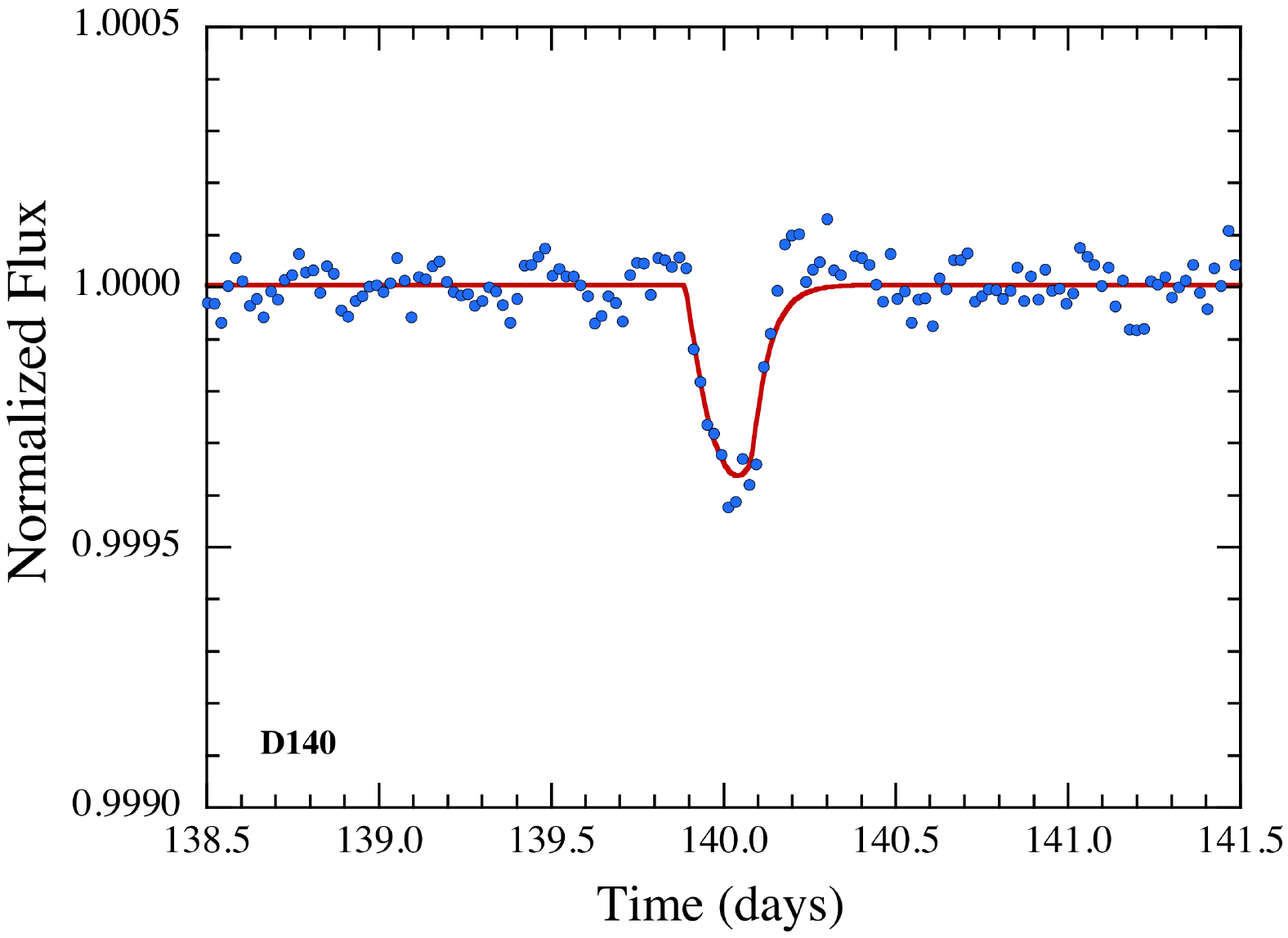} \hglue0.01cm
   \includegraphics[width=2.1in]{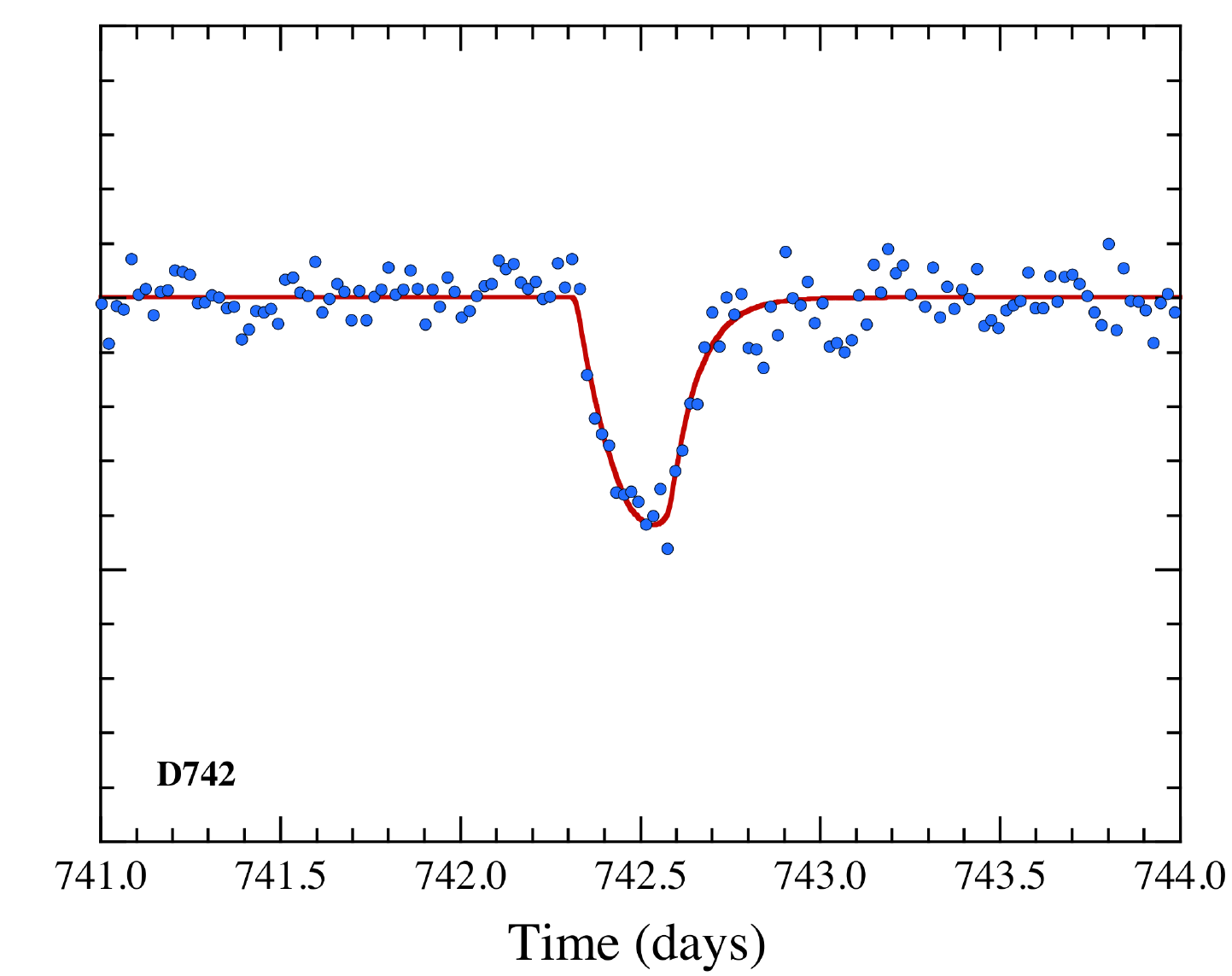} \hglue0.01cm
   \includegraphics[width=2.1in]{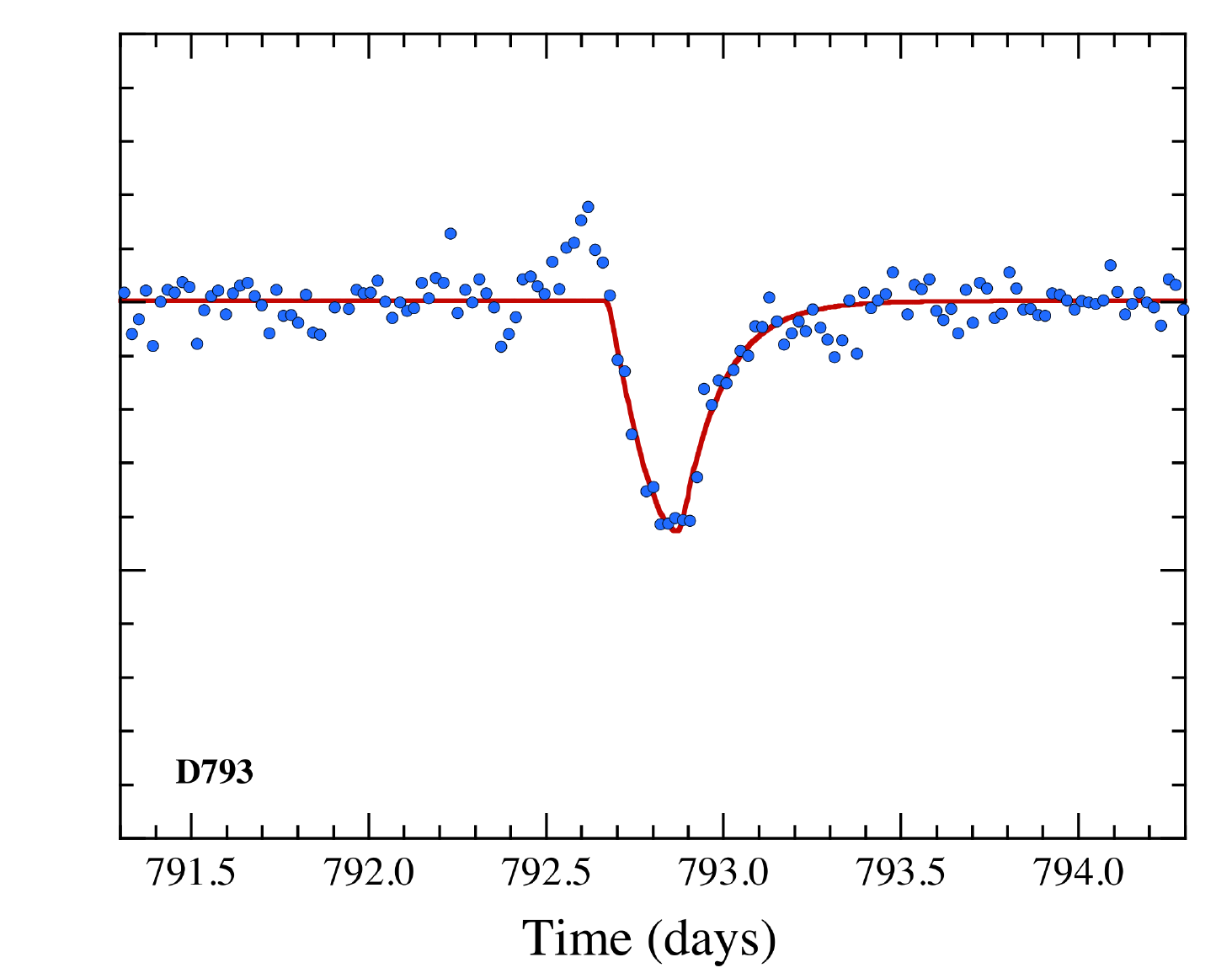} \vglue0.1cm
   \caption{{\em Kepler} SAP photometry covering 3 days around each of the three smaller comet transits.  Other specifications are the same as for Fig.~\ref{fig:transit_profiles}. Note that the vertical (flux) scale has been expanded by a factor of 2 compared to that of Fig.~\ref{fig:transit_profiles}. }
   \label{fig:transit_profiles_small}
\end{figure*}

The three shallower transits we have identified are shown in Fig.~\ref{fig:transit_profiles_small}. Although these dips have lower signal-to-noise, they all appear to have shapes consistent with the asymmetry that is more clearly evident in the deeper, higher signal-to-noise events.  We have ignored all dip-like features whose depth was less than $\sim$450 ppm because of the possibility of having substantial distortions from the spot modulations.  

We tentatively interpret the transits shown in Figs.~\ref{fig:transit_profiles} and \ref{fig:transit_profiles_small} as being due to the passage of comet tails across the disk of the host star, KIC 3542116, as viewed from the direction of the Earth. In this work we henceforth refer to these as `comet transits' and endeavour to demonstrate that they are indeed consistent with the hypothesis of transiting comet dust tails.

\begin{deluxetable*}{lcccccc}
\tablewidth{0pt}
\tablehead{
\colhead{Parameter} & \colhead{Dip 140} & \colhead{Dip 742} &\colhead{Dip 793} &\colhead{Dip 992} & \colhead{Dip 1176} & \colhead{Dip 1268}}
\startdata
 1. Depth (ppm) &$ 491 \pm 38$  & $524 \pm 58$ & $679 \pm 125$ & $1200 \pm 100$ & $1500 \pm 130$ & $1900 \pm 150$ \\
 2a. $v_t^{(a)}$ ($R_*/$day) & $7.76 \pm 0.31$  & $6.55 \pm 0.73$ & $7.42 \pm 0.42$ & $3.04 \pm 0.16$ & $4.34 \pm 0.39$ & $3.70 \pm 0.20$\\
 2b. $v_t$ (km s$^{-1}$) & $89.8 \pm 3.6$  & $75.8 \pm 8.5$ & $85.9 \pm 4.9$ & $35.2 \pm 1.8$ & $50.2 \pm 4.5$ & $42.8 \pm 2.3$ \\
 3. $\lambda^{(b)}$ ($R_*$) & $0.44 \pm 0.04$ & $0.53 \pm 0.09$ & $0.85 \pm 0.16$ & $0.59 \pm 0.10$ & $0.76 \pm 0.11$ & $0.72 \pm 0.08$\\
 4. $b^{(c)}$ ($R_*$) & $0.66 \pm 0.05$ & $0.47 \pm 0.18$ & $0.63 \pm 0.14$ & $0.27 \pm 0.13$ & $0.44 \pm 0.17$ & $0.27 \pm 0.14$ \\
 5. $t_0^{(d)}$ & $139.98 \pm 0.02$ & $742.45 \pm 0.02$ & $792.78 \pm 0.02$ & $991.95 \pm 0.02$ & $1175.62 \pm 0.02$ & $1268.10 \pm 0.02$ \\
\enddata
\tablecomments{a.~Transverse comet speed during the transit; b.~Exponential tail length from Eqn.~(\ref{eqn:attenuation}); c.~Impact parameter; d.~Time when the comet passes the center of the stellar disk.} 
\label{tbl:parms}
\end{deluxetable*}

\section{Assessment and Checks on the Transit Data}
\label{sec:validation}

We performed a set of validation checks on these transit-like events to establish their astrophysical nature and their likely source. These tests included assessing the difference images, analyzing potential video crosstalk (van Cleve \& Caldwell 2016), and inspecting the data quality flags associated with these events.

\begin{figure*} 
   \centering
   \includegraphics[width=\textwidth]{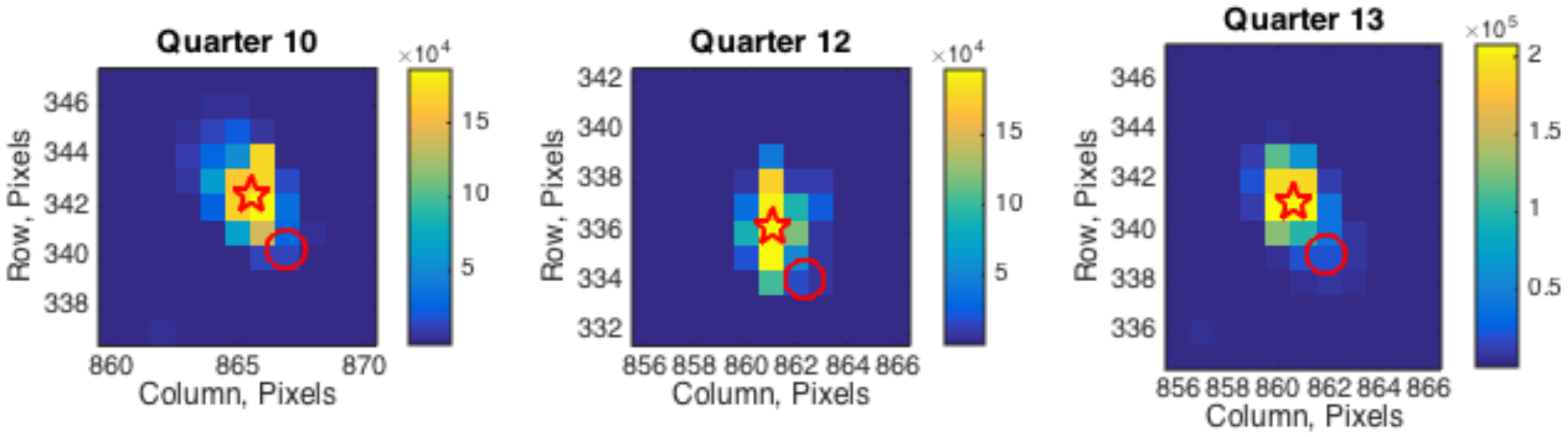} 
   \includegraphics[width=\textwidth]{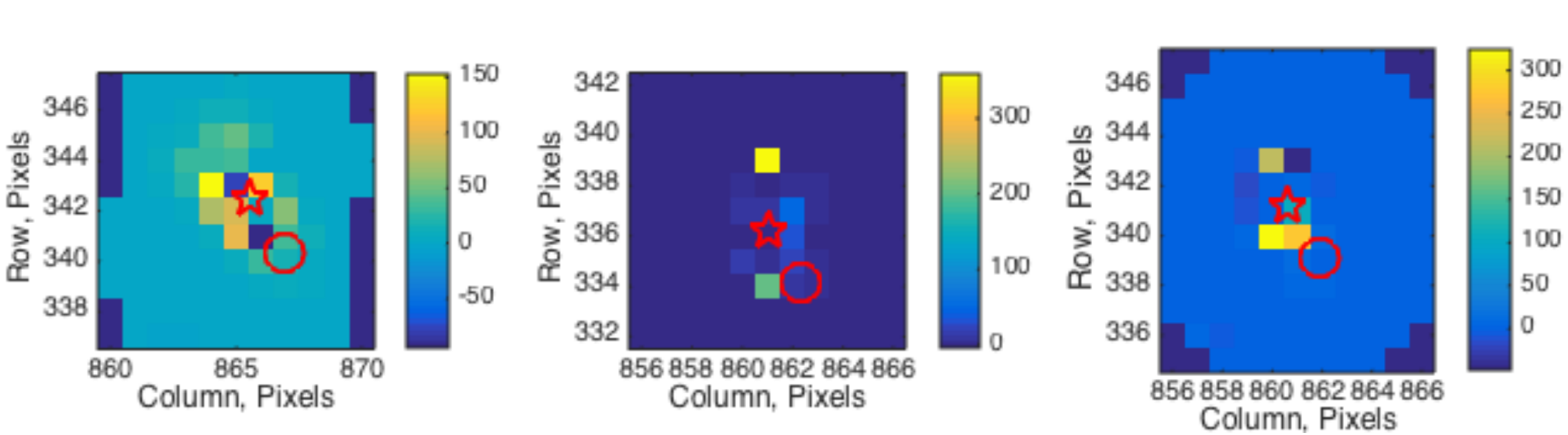}
   \caption{Direct images and difference images for KIC 3542116 during each of the three major transit features for each quarter in which they occur. Top panels show the mean calibrated pixel values in the aperture masks returned by {\em Kepler} for Q10, Q12 and Q13, from left to right. Bottom panels show the mean difference between the calibrated pixels in transit-feature-wide segments on either side of each transit, and the pixel values during each transit feature. The source of the transit features will exhibit positive values in these difference images. The colour-bars indicate the pixel fluxes in units of e$^-$/s. Note that the pixels exhibiting the strongest positive deviations in the bottom panels occur primarily at the ends of the saturated and bleeding columns and are approximately clustered about the location of KIC 3542116, which is marked by a red star.  KIC 3542117's location is marked by a red circle in each panel. These difference images indicate that the source of the transit features is co-located with KIC3542116 to within the resolution of the {\em Kepler} data. }  
   \label{fig:difference_images}
\end{figure*}

To determine the location of the source of the transit signatures, we inspected the pixels downlinked with KIC 3542116 for the quarters containing the three deep events, namely quarters 10, 12, and 13. Since this star is saturated and `bleeding' due to its bright \Kepler\ band magnitude $K_p$ = 9.98\footnote{Stars observed by {\em Kepler} saturate at a magnitude of $\sim$11.5.}, the standard difference image centroiding approach as per \citet{bryson2013} is problematic: small changes in flux can affect the nature of the bleed of the saturated charge and induce light centroid shifts, especially along columns. Indeed, a shift in the flux weighted centroids in the column direction does occur during the Q12 transit, but the direction of the shift is away from KIC 3542116 and toward KIC 3542117, the dim  $K_p \simeq 15$ M-dwarf discussed in Section \ref{sec:discovery} located $\sim$9.8$''$ away from KIC 3542116. This shift is incompatible with the source being KIC 3542117 as the direction is consistent with KIC 3542116 being the source. Figure~\ref{fig:difference_images} shows the direct images of KIC 3542116 and the mean difference image between out-of-transit data and in-transit data, along with the locations of KIC 3542116 and KIC 3542117. Inspection of the pixel time series over the data segments containing the transits reveals that the transit signatures are occurring in the pixels in the core of KIC 3542116 and at the ends of the columns where saturation and `bleed' are happening. While the location of the source of the dips cannot be determined with great accuracy due to the saturation and bleeding, the fact that the transit signatures are not apparent in the saturated pixels but are visible in the pixels just above and below the saturated pixels is strong evidence that the source of the transits is in fact co-located on the sky with KIC 3542116.

As a further check on the astrophysical nature of these events, we also checked against video crosstalk. The \Kepler\ CCD readout electronics do ``talk'' to one another so that dim images (and sometimes negative images) of stars read out on the adjacent three CCD readout channels are electronically superimposed on the image data being read out by the fourth CCD channel \citep{KIH2016}. 

We looked for stars located on the {\em Kepler} detectors which might be the source of any video cross-talk signals by inspecting the full frame images (FFIs) for the quarters during which the three most prominent transits occurred. There is a fairly  bright, possibly saturated star near the edge of the optimal aperture on output 3 on the CCD on which 3542116 is imaged (it's on output 1 in all cases), but the crosstalk coefficient is even smaller than for the other two outputs, $-0.00001$, so that a 50\% deep eclipse on this other star would be attenuated to a value of 5 ppm when its video ghost image is added to the direct image of KIC 3542116, assuming they are the same brightness. Furthermore, since the coefficient is negative, there would need to be a brightening event on the star on output 3 to cause a transit-like dip on output 1.

Fortunately, the largest crosstalk coefficient to the CCD output that 3542116 finds itself on is $+0.00029$, so given that the signal we are looking at is $\sim$0.1\%, a contaminating star would need to be at least 10$\times$ brighter than 3542116 to cause a problem. If there were, it would be highly saturated and bleeding, which would make it difficult to square with the pixel-level analysis indicating that the source is associated with the pixels under 3542116, as the extent of the bleeding would be significantly larger than for KIC 3542116.
        
We also inspected the quality flags associated with the flux and pixel time series for KIC 3542116 and find that the situation is nominal with flags for occasional events such as cosmic rays and reaction wheel desaturations, but no flags for rolling band noise during the transit events.\footnote{See \citet{KAM2016} for more information about anomalies flagged in the {\em Kepler} pixel and flux time series.} 

Finally, we considered `Sudden Pixel Sensitivity Dropouts' (SPSDs) in the data, which are due to radiation damage from cosmic ray hits on the CCD, as a possible explanation for the dips in flux that we observe.  However, the shape and behaviour of such dips do not resemble what we see \citep{KAM2016}.  In particular, the SPSD events have drops that are essentially instantaneous, and therefore are much shorter than the $\sim$20 and 8 long-cadence points on the ingresses that we see in the deeper and more shallow dips, respectively.  Moreover, the location of the SPSDs on the CCD chip would not plausibly align with the source location and its bleed tracks for each and every one of the dips. Thus, we also discarded this idea as well.

We take all these evaluations as strong evidence that the dips we see are of astrophysical origin and that KIC 3542116 is indeed the source of them.  {\em However, we cannot categorically rule out the possibility that the dips are caused by some unknown peculiar type of stellar variability in KIC 3542116 itself.}  In spite of this caveat, we proceed under the assumption that the dips in flux are indeed due to the passage of objects in Keplerian orbit that are trailing tails of dusty effluents.

\section{Ground-Based Studies of KIC 3542116}
\label{sec:groundbased}

The photometric properties of KIC 3542116 are summarised in Table \ref{tbl:mags}. Fortunately, this is a relatively bright star that is amenable to follow-up ground-based study.

\subsection{UKIRT Image}

\begin{figure} 
   \centering
   \includegraphics[width=0.99 \columnwidth]{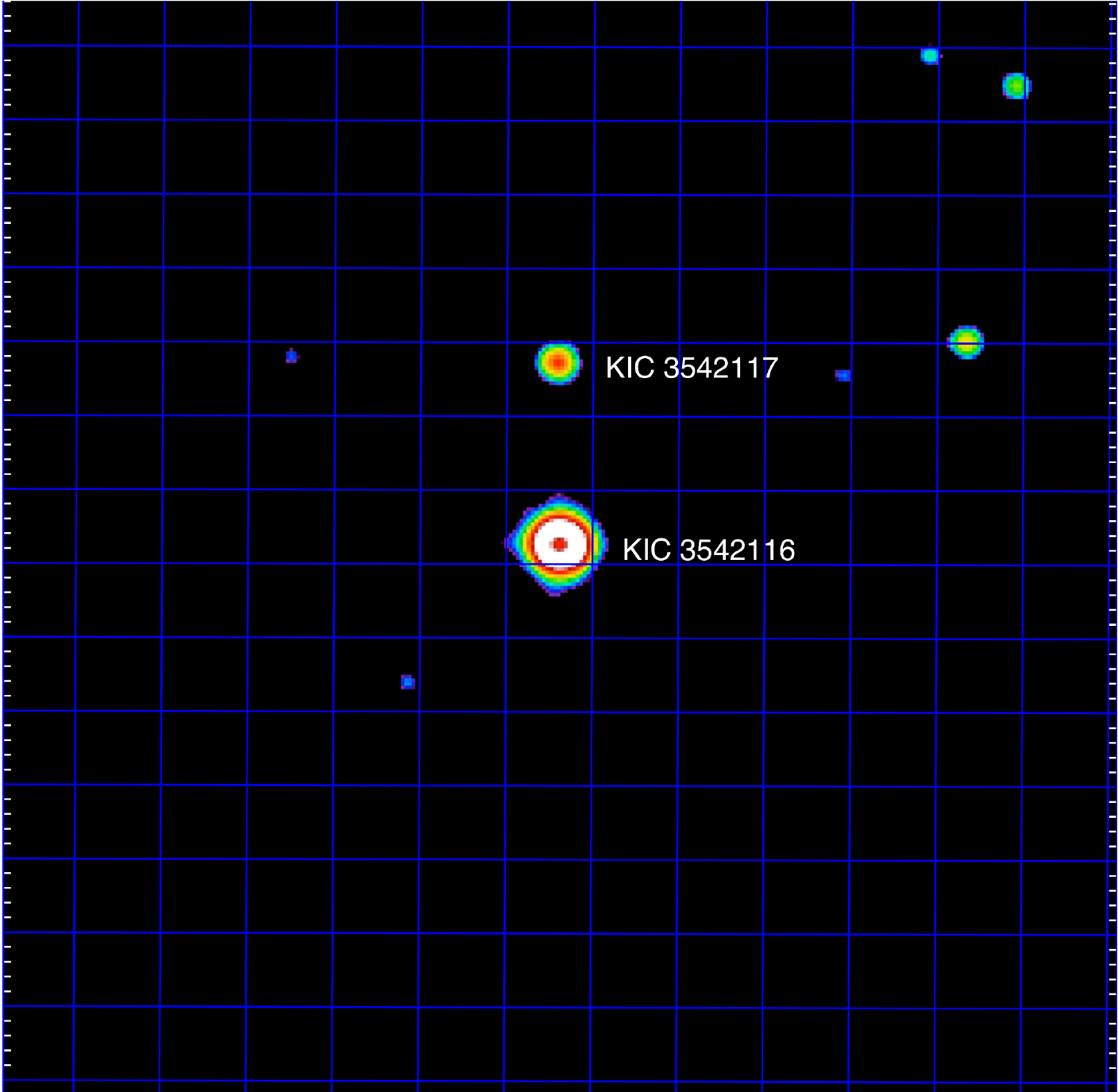}
   \caption{UKIRT J-band image of KIC 3542116.  The colours represent a logarithmic scaling.  Note that the centre of the KIC 3542116 image is saturated.  The grid lines are spaced at $4'' \times 4''$ to match the {\em Kepler} pixels (North is up and East is to the left).  The neighbouring star, KIC 3542117, is $10''$ to the north and has Kp = 14.9 (J = 12.4), compared to Kp = 9.98 (J = 9.3) for KIC 3542116.  The faintest blue-coloured stellar images have J $\simeq 16.0$.}
   \label{fig:UKIRT}
\end{figure}

The UKIRT image of KIC 3542116 is shown in Fig.~\ref{fig:UKIRT}. In addition to the bright target star KIC 3542116 at $K_p = 9.98$, the image shows a neighbouring star, KIC 3542117, with $K_p = 14.9$ some $10''$ to the north. This star is the source of the 23-day modulations (see McQuillan et al.~2014) that leak into the flux data train of KIC 3542116, and may be a low-mass bound companion to this star (see Sect.~\ref{sec:AO}).

\subsection{TRES Classification Spectrum}
\label{sec:TRES}

We observed \thisstar\ with the Tillinghast Reflector Echelle Spectrograph (TRES) on the 1.5m telescope at Fred L.~Whipple Observatory (FLWO) on Mt. Hopkins, AZ. We obtained two high-resolution ($\lambda/\Delta\lambda$ = 44,000) optical spectra of \thisstar\ -- the first on 9 June 2017 and the second on 14 June 2017. Exposure times of 300\,s and 200\,s yielded signal-to-noise ratios of 50 and 43 per resolution element at 520 nm. We cross-correlated the two spectra with a suite of synthetic stellar template spectra from a library of synthetic spectra generated from Kurucz (1992) model atmospheres. These cross-correlations yielded an absolute radial velocity of $-21.1$ km s$^{-1}$. Cross correlating the two spectra against one another and averaging the results over many different echelle orders yielded a shift of only 400 \ms\ between the two spectra.  This is consistent with the photon-limited uncertainties for an F star with a rotational broadening of 57 km s$^{-1}$ (at this SNR the precision would be at least an order of magnitude better for a slowly rotating solar-type star).

\begin{figure} 
   \centering
   \includegraphics[width=0.90 \columnwidth]{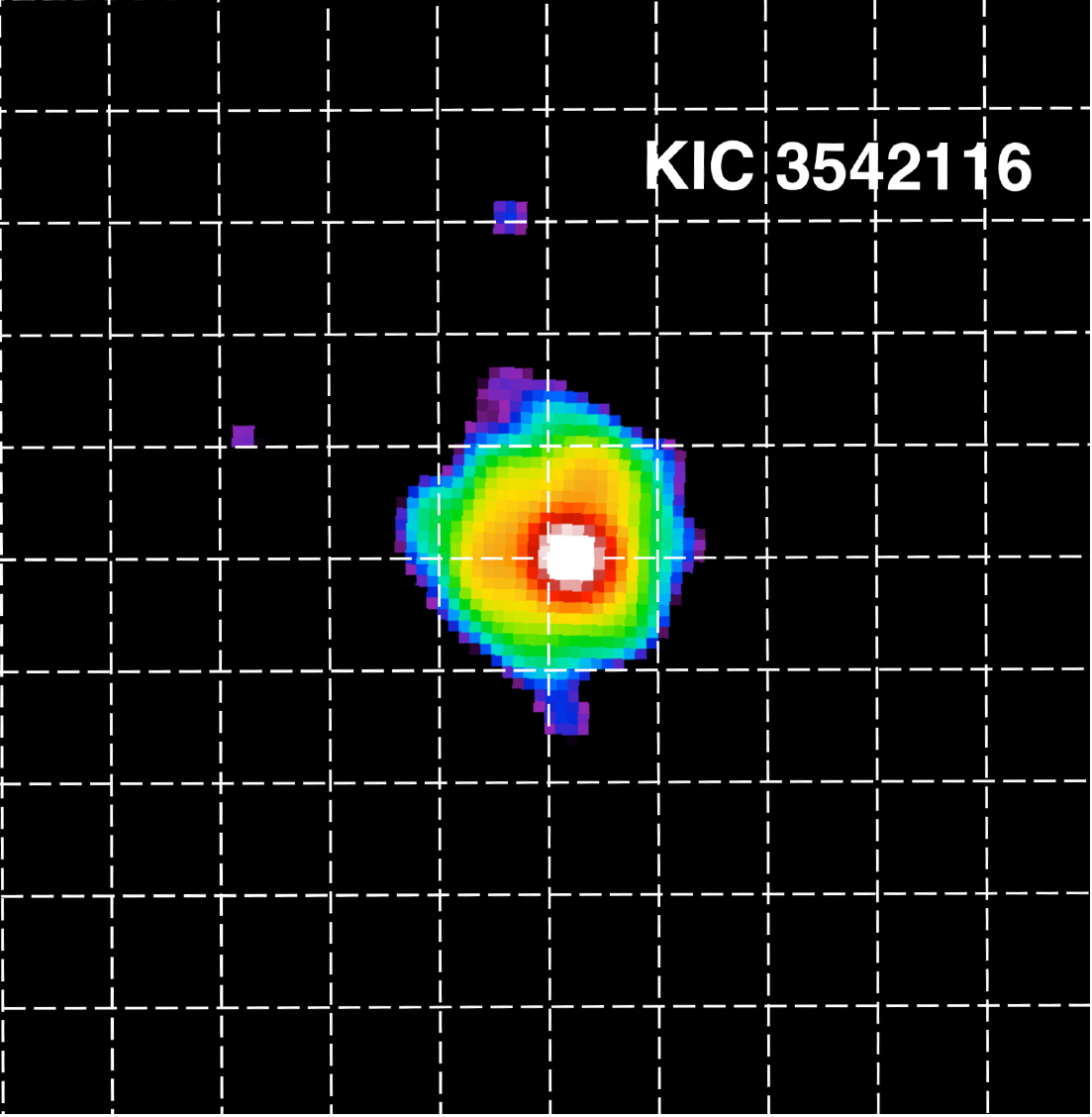} \vglue0.1cm
   \includegraphics[width=0.90 \columnwidth]{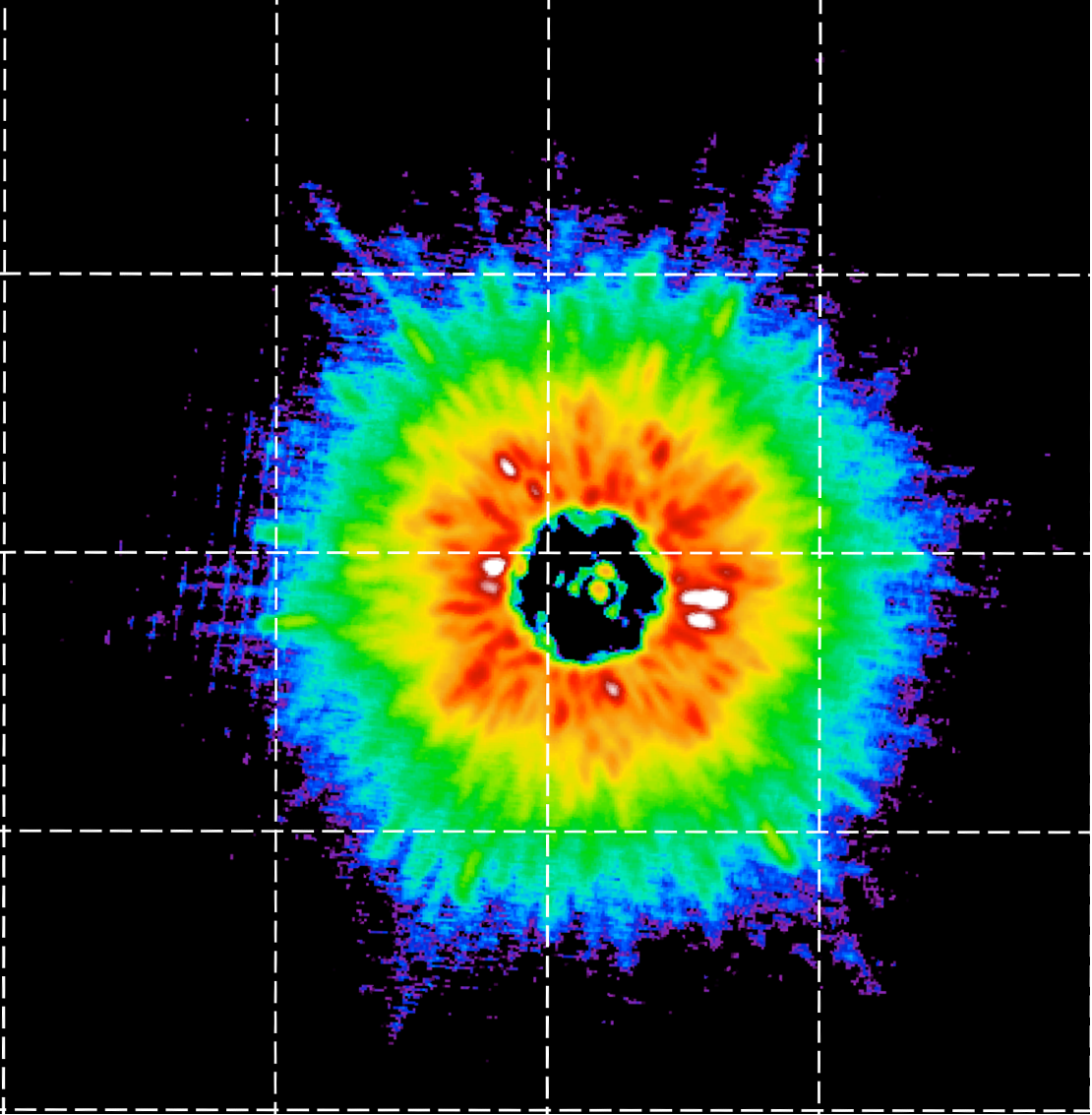}   
\caption{Images of KIC 3542116 from Keck-II/NIRC2 and natural guide star adaptive optics, obtained without (top) and with (bottom) the 0.6$''$-diameter coronagraphic spot. In both panels the pseudocolour is a logarithmic scaling such that each colour difference represents a $\sim$1 magnitude change in brightness. The grid scale in the top (AO) image is $0.1'' \times 0.1''$, while the grid in the bottom panel is $1'' \times 1''$.  The attenuation from the coronagraphic spot ($\Delta K = 7.25 \pm 0.10$ mag; Kraus et al. 2016) allows for unsaturated observations with longer individual exposures and more Fowler samples, which reduces read-noise and allows the detection of fainter sources at wide separations. We see no neighbouring sources out to an azimuthally complete projected separation of $\rho = 4.5$\arcsec, with partial azimuthal coverage (from the corners of the detector) out to $\rho = 7$\arcsec. The wide-separation contrast limits are $\Delta K' = 8.6$ mag and $\Delta K' = 10.8$ mag,
respectively.}
   \label{fig:AO}
\end{figure}  

We estimated stellar parameters using the Stellar Parameter Classification code (`SPC', Buchhave 2012). SPC cross correlates a library of synthetic template spectra with varying temperature, metallicity, surface gravity, and line broadening against the observed spectrum and interpolates the parameters from the best-matched template peaks to estimate the actual stellar parameters. SPC was designed to measure stellar parameters of slowly rotating stars close in effective temperature to the Sun, and has been extensively tested and used for stars cooler than the Sun.  For rapidly rotating stars hotter than the Sun (such as KIC 3542216), SPC has not been tested as fully and may have systematic errors, especially in the surface gravity and metallicity.  

An SPC analysis of the TRES spectra of KIC 3542116 yields an effective temperature of $6900 \pm 120$ K and a projected rotational velocity of 57 \kms.  This makes it fairly unusual for stars observed by \Kepler, which mostly observed sun-like dwarfs and smaller stars, which are more amenable for searches for small Earth-like planets. The properties of KIC 3542116 measured with, or inferred from, the TRES spectra are summarised in Table \ref{tbl:mags}.

\subsection{High-Resolution Imaging}
\label{sec:AO}

We observed \thisstar\ with the Near Infrared Camera 2 (NIRC-2) instrument behind the Natural Guide Star (NGS) adaptive optics system on the Keck II telescope on the night of 28 June 2017. We obtained standard adaptive optics (`AO') images, both with and without a coronagraph in place.  We also recorded interferograms produced by placing a sparsely sampled nine-hole non-redundant aperture mask (`NRM') in the pupil plane to resample the full telescope aperture into an interferometric array (Tuthill et al.~2006; 2010). This process makes it possible to detect companions closer to the target star than the traditional diffraction limit.   We obtained four 20-s exposures in imaging mode in K$'$ band, as well as four exposures of the same duration with the coronagraph.  Six additional 20-s exposures were taken with the NRM in place.  The imaging observations were reduced following Kraus et al.~(2016), and detections and detection limits were assessed using the methods they described. 

The summed set of the standard AO images is shown in the top panel of Fig.~\ref{fig:AO}; it covers only the central $1'' \times 1''$ region of the field.  The bottom panel in Fig.~\ref{fig:AO} shows the resultant image acquired with the coronagraphic disk in place, and it covers a wider $4'' \times 4''$ portion of the field.  The images are colour coded so that roughly each contrast change of 1 magnitude is represented by a change of one colour. From the ordinary AO image we can estimate that there is no neighbouring star of comparable K$'$ magnitude within 0.05$''$ of the target star, and no star that is at most 4 magnitudes fainter within 0.15$''$.  With the AO-plus-coronagraph image the sensitivities are comparable out to about 0.8$''$, beyond which the image goes 2 magnitudes deeper than the plain AO image.  

We can further constrain the magnitudes of any stars within $\sim$0.2$''$ of the target, using the 9-hole mask interferogram.  The analysis of those data shows that contrasts of $<$ 1.5, 4.2, 5.1, and 4.8 K$'$ magnitudes can be rejected at the 99\% confidence limit at distances of 0.01-0.02$''$, 0.02-0.04$''$, 0.04-0.08$''$, 0.08-0.16$''$, respectively.   

\begin{figure} 
   \centering
   \includegraphics[width=0.99 \columnwidth]{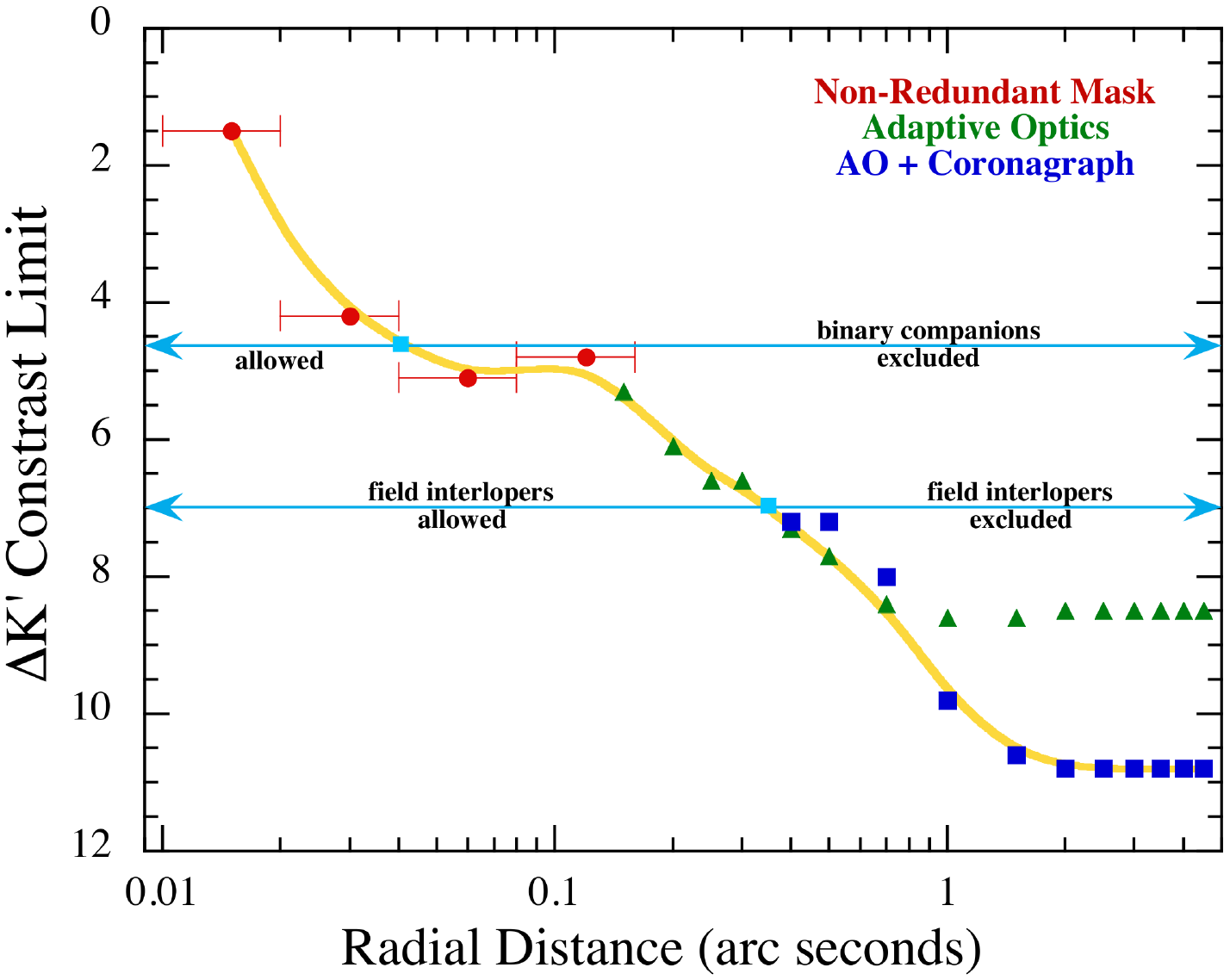}
   \caption{Contrast limits (in magnitudes) for possible neighboring stars to the target star, KIC 3542116, as a function of radial distance.  These were obtained with high resolution imaging using the Keck telescope (see Sect.~\ref{sec:AO}).  The red circles, green triangles, and blue squares are for the non-redundant mask technique, natural guide star adaptive optics, and AO plus a coronagraph (see text for details).   The yellow curve is a smooth fit to the best contrast constraints at any given radial distance. The working region within the field goes out to 4.5$''$.  Neighbouring field stars farther from the target than $\sim$0.4$''$ are marked as ``field interlopers excluded'' because they would have insufficient flux to produce a 0.15\% dip in the optical flux of KIC 3542116 unless they are extremely blue. Physical binary companions with separations $\gtrsim 0.04''$ are labeled ``binary companions excluded'' because they also would have insufficient flux to produce a 0.15\% dip in the optical flux of KIC 3542116. }
   \label{fig:contrast}
\end{figure}

We summarise all of the constraints from the three different imaging modes in Fig.~\ref{fig:contrast}.

We now consider the constraints we can place on possible neighbouring stars in each of two different categories: (i) random interloping field stars, and (ii) physically bound companions.  For unbound field stars, we see from Fig.~\ref{fig:contrast} that for angular separations greater than $\simeq 0.4''$ there are no stars within 7 K$'$ magnitudes of the target star.  The significance of this latter limit is that stars fainter than $K_p = 17$th magnitude\footnote{For neighbouring field stars cooler than $T_{\rm eff} \simeq 7000$ K the contrast limits in $K_p$ are even more stringent than in K$'$.  However, we estimate that the contrast constraints obtained in K$'$ are still good to within $\simeq 1$ magnitude in $K_p$ for $T_{\rm eff}$ of the hypothetical interloper star up to $\simeq 15,000$ K.  For even higher $T_{\rm eff}$ the constraints weaken further only very slowly.} could not produce a dip as apparently deep as 0.0015 in the flux of KIC 3542116.  It is possible, though most unlikely, that there could be a star accidentally aligned with the target to within $\lesssim 0.4''$ that is the source of the dips.  We estimate the probability of a nearby interloper star with the requisite magnitude of $ K_p \lesssim 17$ randomly lying within 0.4$''$ of \thisstar\ as $\lesssim 0.1\%$ (see, e.g., Fig.~9 of Rappaport et al.~2014b).

Alternatively, the target star could have a physical binary companion that is the host of the dips.  In this  case, the companion star would be at the same distance and coeval with the target star.  Since the target star is not significantly evolved (see Table \ref{tbl:mags}) any fainter companion star would necessarily be redder in colour and lower in mass than the target star.  For redder stars, the K$'$ band is obviously more sensitive than the $K_p$ band.  From Fig.~\ref{fig:contrast} we see that for any binary (projected) separation $\gtrsim 0.04''$ all binary companions with $\Delta K' \gtrsim 4$ are ruled out.  That already suggests that  any companion-star mass satisfying this requirement must be $\lesssim 0.8 \, M_\odot$.  However, for main-sequence stars of this mass, and lower, the value of $\Delta K_p$ to be expected in the {\em Kepler} band would be 2-3 magnitudes greater.  Thus, it is safe to say that for angular separations $\gtrsim 0.04''$ there are no binary companion stars that could produce the observed dips.  This translates to a binary orbital projected separation of $\lesssim 10$ AU.  If we combine this with the constraint on the change in RVs over a 5-day interval (Sect.~\ref{sec:TRES}), this suggests that any viable binary companion star that could produce the dips would likely have an orbital separation of $\sim$0.5-10 AU.

Of course very faint binary companions that are {\em not} the source of the dips are allowed.  They must, however, still satisfy the constraints summarised in Fig.~\ref{fig:contrast}. In this regard, we note that the faint star, KIC 3542117 (see Fig.~\ref{fig:UKIRT}) has the colours (taken from the Sloan Digital Sky Survey images; Ahn et al.~2012) to be a $\simeq 0.5 \, M_\odot$ (see also Dressing \& Charbonneau 2013) companion star since it lies relatively close to the same photometric distance as KIC 3542116, and because it has only a $\lesssim 15\%$ chance of being found within 10$''$ of the target star by chance (see, e.g., Fig.~9 of Rappaport et al.~2014b).  We note, however, that the proper motion of KIC 3542117 is not consistent with comovement, though the Deacon et al.~(2016) seeing-limited astrometry might be limited by the brightness of KIC 3542116.  Any possible association should become clear in the upcoming Gaia Data Release 2.

\subsection{Historical Plate Stacks}

\begin{figure} 
   \centering
   \includegraphics[width=0.99 \columnwidth]{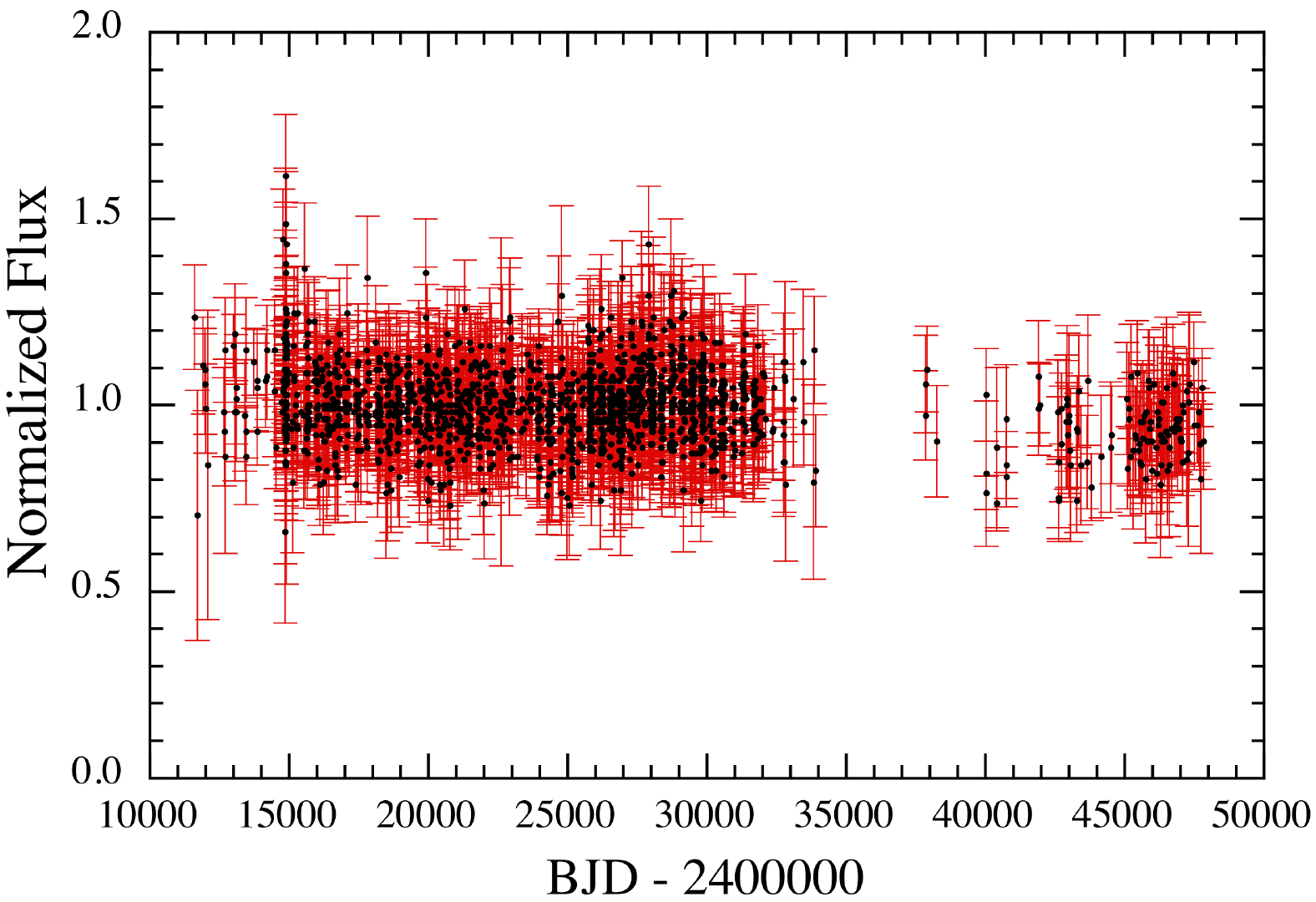}
   \caption{One hundred years of photometry on KIC 3542116 from the Harvard Plate Stack collection, ``Digital Access Sky Century at Harvard'' (`DASCH'; Grindlay et al.~2009). The systematic drop in flux, by $\sim$10\% across the `Menzel gap', is likely due to a change in the plate emulsion response.}
   \label{fig:DASCH}
\end{figure}

The available photometry of KIC 3542116 from the past century, taken from the ``Digital Access Sky Century at Harvard'' (`DASCH'; Grindlay et al.~2009) are shown in Fig.~\ref{fig:DASCH}.  The systematic drop in flux, by $\sim$10\% across the `Menzel gap', is likely due to a change in the plate emulsion response.  No other obvious dimmings or outbursts of the star are observed. A Fourier transform of these data show no clear periodicities in the range of $1-100$ days over the past 100 years down to a level of $\gtrsim$ 2\%.  

\vspace{0.8cm}

\section{Model Fits to the Transits}
\label{sec:model}

The repeatably asymmetric shape of the transits of \thisstar\ is suggestive of an occulter with some sort of sustained or repeatable dust outflows causing the dimming. Here, we show that the observed asymmetry and the repeatable shape are consistent with, and what we might expect from, a large comet transiting the star with a dusty tail.  We describe a simple model for the transit of a comet and show that the six transits detected in the lightcurve of \thisstar\ are well fit by this model. 

Almost all cometary dust tails will lag behind the direction in which the comet is moving.  This is as opposed to the ion tails of comets which are driven out nearly radially by the stellar wind of the host star (see, e.g., Reyes-Ruiz et al.~2010). When the dust is released from the immediate and gravitational environs of the comet, it finds itself in a reduced effective gravity due to the effects of radiation pressure on the dust grains. This, in turn, results in the dust moving too fast to remain in the same orbit as the parent comet. Somewhat paradoxically, the higher speed causes the dust to go into a higher orbit which, in turn, causes its mean orbital speed to decrease, and thus results in a trailing tail (in the sense of lagging in angular position).

Leading dust tails are also possible, but usually in the context of the dust overflowing the Roche lobe of the parent body, and with little subsequent radiation pressure (see, e.g., Sanchis-Ojeda et al.~2016). However, Roche-lobe overflow typically requires orbital periods of $\lesssim 1$ day.

In this work we assume that the orbital periods of the putative comets are substantially longer than the transit durations ($\sim$1 day), and that the transverse velocity, $v_t$, during the transit is essentially a constant.  For lack of more detailed information, we further assume that the dust tail is narrow compared to the size of the host star, and that its dust extinction profile is a simple exponential function of distance from the comet (see, e.g., Brogi et al.~2012; Sanchis-Ojeda et al.~2015).

We take the comet location, projected on the plane of the sky, to be $\{x_c,y_c\}$, and an arbitrary location on the disk of the star to be $\{x,y\}$, where $\{0,0\}$ is the center of the stellar disk, and $x$ is the direction of motion of the comet.  We model the attenuation of starlight at $\{x,y\}$, as
\begin{eqnarray}
\tau & = & C \, e^{-(x_c-x)/\lambda} ~~{\rm for}~x < x_c; ~|y-b| \le \Delta b/2 \label{eqn:attenuation} \\
\tau & = & 0 ~~{\rm otherwise}
\end{eqnarray}
where $C$ is a normalisation constant equal to the optical depth of the dust tail just behind the comet, $\lambda$ is the exponential scale length of the tail, $b$ is the impact parameter of the transit, and $\Delta b$ is the width of the dust tail as projected on the disk of the host star\footnote{$\Delta b$ is an undetermined parameter that is essentially degenerate with the normalisation constant $C$}. We assume that $\Delta b \ll R_*$.  The comet position along the $x$ direction is given as a function of time, $t$, by:
\begin{equation}
x_c = v_t (t-t_0)
\end{equation}
where $t_0$ is the time when the comet crosses the centre of the stellar disk. We also adopt a quadratic limb darkening law for the host star, with coefficients appropriate to a mid-F star (Claret \& Bloemen 2011).  

We utilize a Markov Chain Monte Carlo (`MCMC') code (Ford 2005; Madhusudhan \& Winn 2009; Rappaport et al.~2017) to fit a comet-tail transit model to each of the six transits that we observe in KIC 3542116. There are six free parameters to be fit: $t_0$, $\lambda$, $C$, $v_t$, $b$, and DC, where the final term is the DC background flux level away from the transit. For each choice of parameters, we generate a model lightcurve by integrating over the dust tail where it overlaps the stellar disk, and repeating this in increments of 6 minutes as the comet crosses the stellar disk. The model lightcurve is then convolved with the 30 minute integration time of the {\em Kepler} long-cadence sampling. Each model is then evaluated via the $\chi^2$ value of the fit to the data, and the code then decides, via the Metropolis-Hastings jump condition, whether to accept the new set of parameters or to try again.

\begin{figure*} 
   \centering
   \includegraphics[width=3.0in]{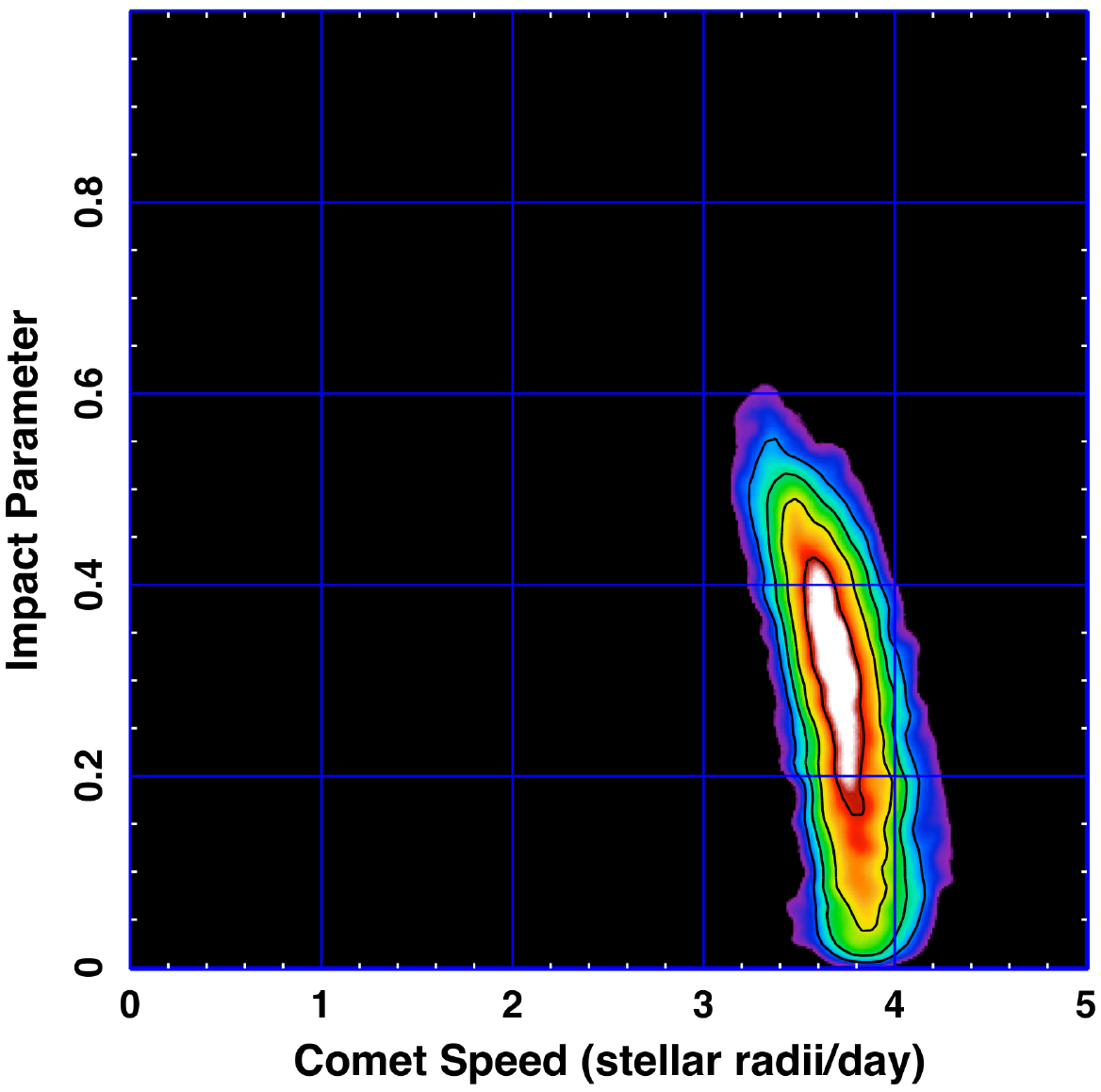} \hglue0.1cm
   \includegraphics[width=3.0in]{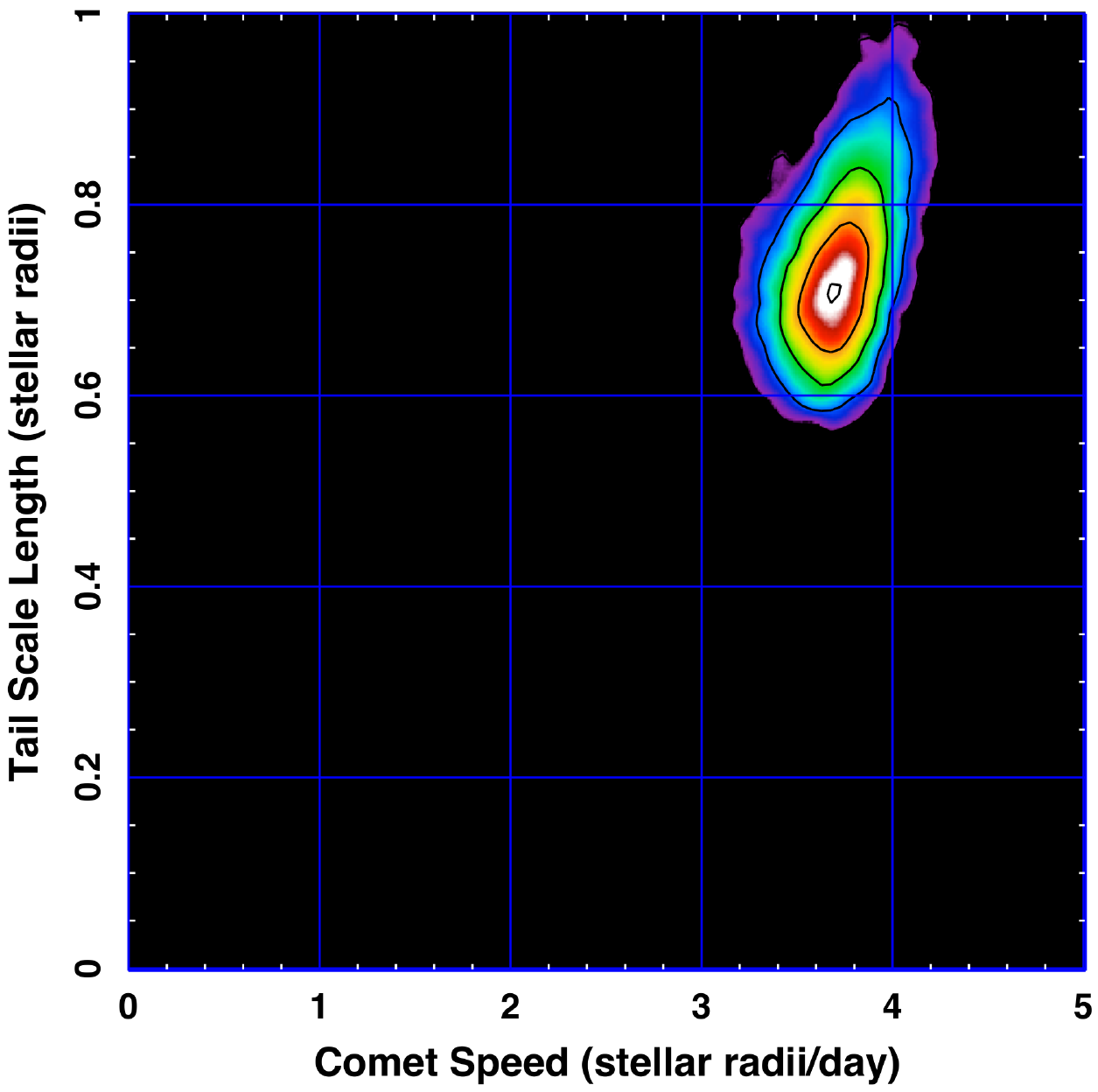} 
   \caption{Illustrative examples of the correlations among the MCMC fitted parameters for the comet speed, $v_t$, impact parameter, $b$, and exponential tail length, $\lambda$ (see Sect \ref{sec:model}). This particular fit was for transit D1268. The colour scaling is logarithmic.}
   \label{fig:MCMC_parms}
\end{figure*} 

Each MCMC chain has $3 \times 10^5$ links, and we run a half-dozen chains.  The results for the fitted parameters are given in Table \ref{tbl:parms} for each of the parameters and for each of the six transits. We show illustrative MCMC correlations among the three physically interesting parameters, $\lambda$, $v_t$, and $b$, graphically for the fit to transit D1268 in Fig.~\ref{fig:MCMC_parms}. 

\vspace{0.3cm} 
\section{Interpreting the Transits}
\label{sec:interpret}

Now that we have shown that the shapes of the transits of \thisstar\ are consistent with the shape caused by a dusty tailed comet, in this section, we will show that the parameters we derive from our fits correspond to plausible physical conditions. 

\subsection{Inferred Comet Orbital Velocities}
\label{sec:orbits}

The first step in trying to understand what orbits the putative comets orbiting KIC 3542116 would be on, is to attempt to explain the observed transverse speeds of the bodies.  This involves speeds of $35-50$ km s$^{-1}$ for the deeper transits and $75-90$ km s$^{-1}$ for the more narrow and shallow transits (see Table \ref{tbl:parms}). To gain some insight into this problem, we carried out the following exercise.  We chose random orbital periods from a distribution that is uniform in log $P_{\rm orb}$, and with a uniform distribution of eccentricities from $e = 0$ to 1. With respect to a fixed viewing direction, we also chose longitudes of periastron, $\omega$, at random from 0 to 2$\pi$ ($\omega$ is here defined as the angle from entering the plane of the sky to periastron). For simplicity, all orbits are taken to be in the same plane with an inclination angle of 90$^\circ$ with respect to the observer.

\begin{figure*} 
   \centering
   \includegraphics[width=3.0in]{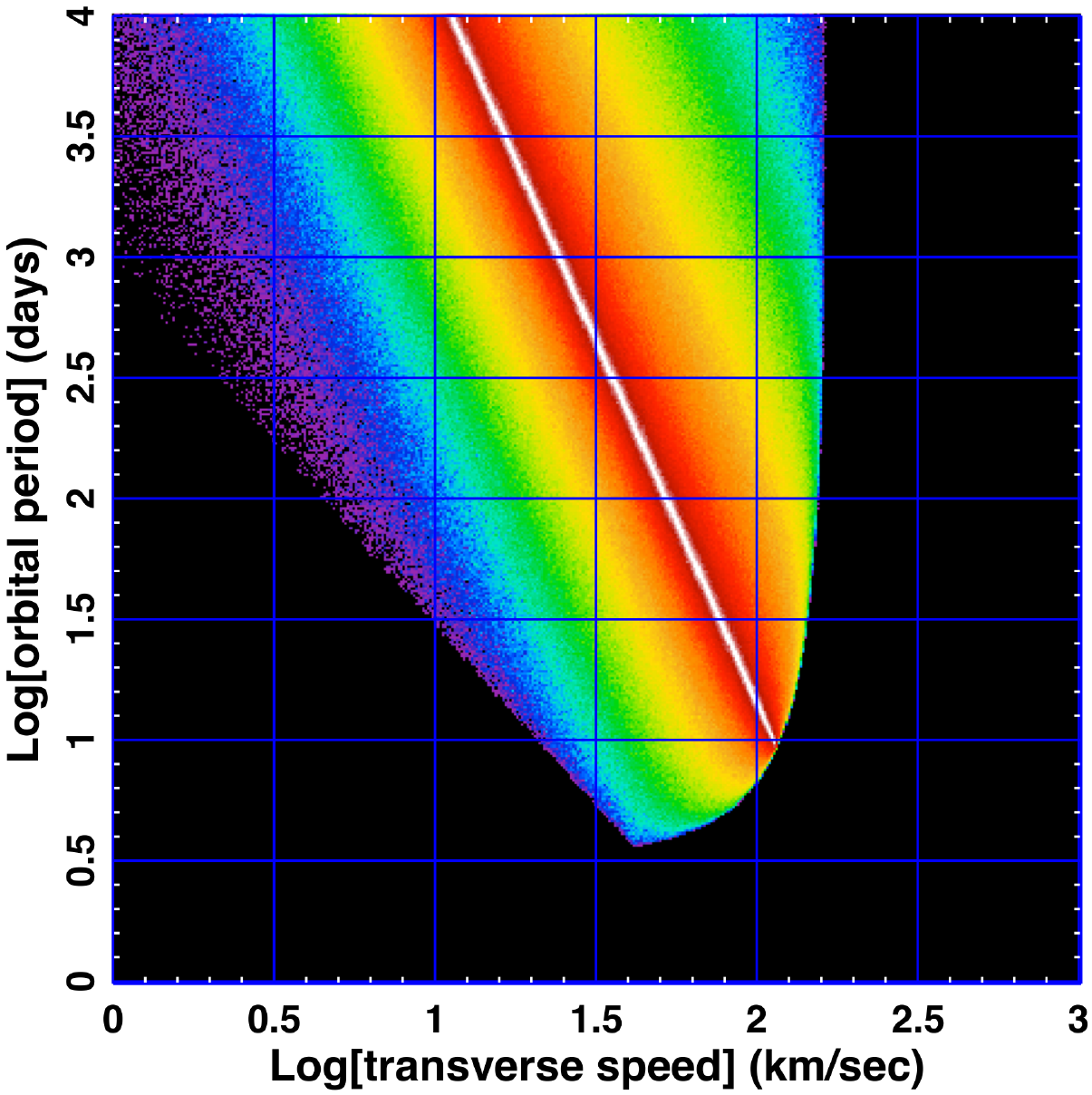}   \hglue0.1cm
   \includegraphics[width=3.0in]{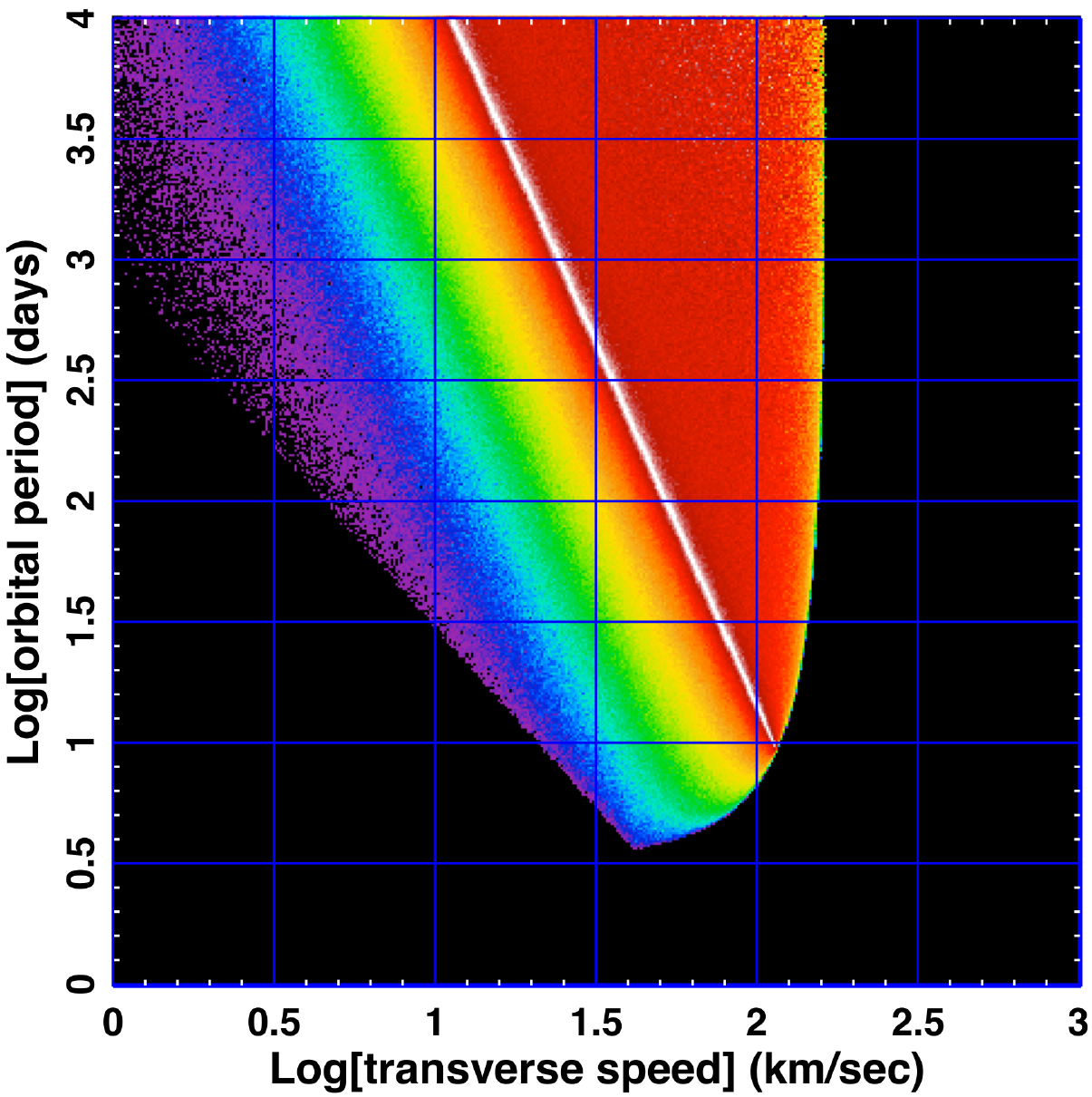}  
   \caption{Statistical assessment of different comet orbital periods and eccentricities.  The orbital periods are chosen randomly from a uniform distribution per logarithmic interval, while the eccentricities and arguments of periastron were chosen randomly between 0 and 1, and 0 and 2$\pi$, respectively.  For each orbit the transverse speed across the disk of the host star was recorded.  In the left panel the weighting for each system is 1.0, while in the right panel, the weighting is proportional to $a/d$, where $a$ is the comet semimajor axis, and $d$ is the star-comet separation during the transit.  All systems where the star-comet separation is $<$ 0.1 AU during the transit are eliminated due to dust sublimation that would destroy the tail (this forms the lower and right boundaries of the diagram). The colour coding is logarithmic (white is highest and purple lowest) and reflects the chosen weighting. For other details see text.}
   \label{fig:MC_p_v}
\end{figure*}

For a Keplerian orbit there is a relatively simple relation among the transverse speed during transit, $v_t$, the orbital period, and the orbital eccentricity
\begin{equation}
v_t = (2 \pi G M)^{1/3}  P_{\rm orb}^{-1/3} [1-e \sin \omega ] (1-e^2)^{-1/2}
\label{eqn:vt_P}
\end{equation}
Also, at the time of transit, the separation between the host star and comet can be written analytically as
\begin{eqnarray}
d  & = & a(1-e^2) [1-e \sin \omega]^{-1}Ê\nonumber \\
& = & (2 \pi)^{-1/3} (GM)^{2/3} P_{\rm orb}^{1/3} \sqrt{1-e^2} \, v_t^{-1}  
\label{eqn:dist}
\end{eqnarray}
After choosing $10^7$ such random orbits we record the mean transverse speed of each body as it transits the host star.  The results are shown in Fig.~\ref{fig:MC_p_v}.  To construct the left panel we simply add the value 1 to each pixel in the $P_{\rm orb}-v_t$ plane where a system appears.  By contrast, in the right-hand panel, the weight given to each system is taken to be proportional to $a/d$, where $a$ is the comet semi-major axis and $d$ is the separation of comet and host star during the transit. The $1/d$ factor is supposed to represent the probability of a transit if the comets {\em were not} all in the same plane, but rather in a range of randomly tilted planes. The extra factor of $a$ is simply to (i) render the weighting factor dimensionless, and (ii) restore some weight to the longer-period systems, so that the shorter period systems don't dominate the diagram.  

As discussed below in Sect.~\ref{sec:sublimation}, comet dust tails are very unlikely to survive sublimation at distances from the host star of $\lesssim 0.1$ AU (see Eqn.~\ref{eqn:dist} and Fig.~\ref{fig:tequil}).  Therefore, any systems with such close approaches during the transit are eliminated from the diagram.  Equations (\ref{eqn:vt_P}) and (\ref{eqn:dist}) can be combined to yield analytic relations for the upper limit to $v_t$ and lower limit to $P_{\rm orb}$ in Fig.~\ref{fig:MC_p_v} for a given minimum allowed star-comet separation during transit: $v_t < 160 \, d_{0.1}^{-1/2}$ km s$^{-1}$ and $P_{\rm orb} > 3.5 \, d_{0.1}^{3/2}$ days, where the subscript on $d$ indicates that it is normalised to 0.1 AU.  These boundaries are clearly evident in Fig.~\ref{fig:MC_p_v}.

The white track in the Fig.~\ref{fig:MC_p_v} diagram is just the locus of points given by
\begin{equation}
P_{\rm orb} \propto  v_t^{-3}
\label{eqn:P_vt}
\end{equation}
that would hold for a simple circular orbit.  Points to the right of the white track are eccentric orbits with the periastron passage on the side of the orbit nearest the observer while systems to the left are eccentric orbits with periastron on the far side of the orbit.  In the right-hand panel of Fig.~\ref{fig:MC_p_v} the higher likelihoods to the right of the white band arise from the inverse weighting with separation between the host and comet during the transit.  The higher speeds reflect closer distances, and therefore higher transit probabilities.    
    
As we can see from Fig.~\ref{fig:MC_p_v}, transverse speeds of $35-50$ km s$^{-1}$ would correspond to circular orbit periods of $\sim$$100-300$ days, but periods as short as $\sim$6 days are plausible. Note, especially from the right-hand panel in Fig.~\ref{fig:MC_p_v}, that arbitrarily long orbital periods are quite possible.  Similarly, the more narrow and shallow dips imply transverse orbital speeds during transit ranging between $\sim$75 and 90 km s$^{-1}$. From Fig.~\ref{fig:MC_p_v} we see that such orbits would  correspond to circular orbit periods of $\sim$20-35 days.  However, periods as short as $\sim$6 days or arbitrarily long are also quite acceptable.

One possibility is that all three of the deeper transits arise from a single body in a periodic orbit (or nearly so).  In that case, the orbital period would be 92 days, but would have to exhibit transit timing variations (`TTVs') of $\sim$1/3 day.  Additionally, there would be the issue of why only three transits appear out of a possible $\sim$16 that potentially could have been detected during the {\em Kepler} main mission. Presumably, such an explanation requires highly and remarkably variable dust emission from the body.  In this regard, we note that the dust-tail optical depths of some Solar-system comets are highly variable (see, e.g., Montalto et al.~2008; Sekanina \& Chodas 2012; Knight \& Schleicher 2015), and that the transit depths in the `disintegrating' planets KIC 1255b and K2-22b are also highly and erratically variable (Rappaport et al.~2012; Sanchis-Ojeda et al.~2015).  If the three deeper transits are indeed due to a single body orbiting with TTVs of up to $\sim$1/3\% of the orbital period, then only the discrete periods of $92/n$, where n = 1, 2, 3, ... are allowed.

Similarly, the three more shallow and narrow transits could be due to another distinct body orbiting KIC 3542116.  The maximum such period consistent with these three transits is $\sim$51 days (the time interval between D742 and D793).  If these three shallower transits are indeed due to a single body in a fixed orbit, then one must explain why only three of a possible 30 transits are detected. Again, this would require highly variable dust emission.

Alternatively, all 6 of the transits could each be due to a separate body in the system.  In that case, one needs to explain why all three of the deeper transits are so remarkably similar in depth, shape, and duration.  And, to a somewhat lesser degree, the same argument applies to the three more shallow transits.  We therefore tentatively adopt the working hypothesis that there are at least two distinct orbiting minor bodies in KIC 3542116 with cometary tails that produce transits.

Finally, in a related vein, we give some estimates of the transit probabilities for the comets we believe we have detected.  Since the transit probability is $R/d$, we can rewrite Eqn.~(\ref{eqn:dist}) as:
\begin{equation}
\frac{R}{d} \simeq 0.0076 \left(\frac{v_t}{40~{\rm km/s}}\right) \left(\frac{P_{\rm orb}}{1~{\rm yr}}\right)^{-1/3}  (1-e^2)^{-1/2}
\end{equation}
where we have normalised the transverse speeds to 40 km s$^{-1}$ which is roughly the mean value for the larger transits. Thus, the transit probability is not particularly large unless either the orbital period is quite short or, more likely, the orbital eccentricity is close to unity.  For example, if $e = 0.99$, the transit probability rises to 5\%.
                                                     
\subsection{Inferred Dust Mass Loss Rates}   
\label{sec:mdot}
                        
 It is straightforward to estimate a lower limit to the instantaneous mass contained in a comet tail that would be required to produce a $\sim$0.1\% transit depth.  Since the most effective visual extinction per unit mass occurs for $\sim$1 $\mu$m particles, consider a dust sheet of thickness, $h$, that is between the observer and the host star which we take to have radius, $R_*$. The minimum projected area of the dust cloud particles that is required to block 0.1\% of the star's light is $\Delta A_d \simeq 0.001 \pi R_*^2$.  The corresponding mass in such a dust sheet is at least:
 \begin{equation}
\Delta M_d \gtrsim 0.001 \pi R_*^2 h \rho_d \simeq 10^{16} ~{\rm g}
 \end{equation}
 where $\rho_d$ is the bulk density of the dust, and where we have taken $\rho_d \simeq 3$ g, and $h = 1 \, \mu$m.
 
 Without knowing the specific properties of the dust or the comet orbit, it is difficult to know the speed of the dusty effluents with respect to the comet. However, if we assume a minimal value for $\beta$, the ratio of radiation pressure to gravity, of $\sim$0.05, the relative dust speed could be $\sim$0.1 times the orbital speed of the comet (see, e.g., Rappaport et al.~2014a), or some 5 km s$^{-1}$.  At this rate, the dust tail at $\sim$2 $R_*$ from the comet would be replenished every $\sim$5 days.  This, in turn, corresponds to a minimum dust mass loss rate of $\dot M_d \gtrsim 2.5 \times 10^{10}$ g s$^{-1}$.
 
 Finally in this regard, if we assume that the comet emits dust at this rate for even half of the interval between the D992 and D1268 transits (276 days), during which time there were three of a possible four transits seen, then the minimum comet mass would be $M_c \gtrsim 3 \times 10^{17}$ g.  This is just a little bit greater than the mass of Halley's comet.
 
 \subsection{Dust Sublimation}
 \label{sec:sublimation}
 
 In order for a dust-rich comet tail to exist it should not be so close to the host star that the dust grains leading to most of the opacity quickly sublimate (i.e., on less than a timescale of about a day).  The equilibrium temperature of the dust, $T_{\rm equil}$, depends mainly on the stellar flux at its location, its size, $s$, and the imaginary part of its index of refraction, $k$, at the wavelengths of interest.  We have computed $T_{\rm equil}$ for three different characteristic grain sizes, $s = 0.1$, 1, and 10 $\mu$m as functions of their distance from KIC 3542116. In lieu of discussing any particular mineral composition for the grain size, we simply adopt three illustrative values for $k$ equal to  1, 0.1 or 0.01, that are taken to be independent of wavelength, and are fairly representative of different refractory minerals (see, e.g., Beust et al.~1998; Fig.~13 of Croll et al.~2015; van Lieshout et al.~2014; Xu et al.~2017; and references therein).  The Mie scattering cross sections were computed with the Boren \& Huffman (1983) code.
 
\begin{figure}
   \centering
   \includegraphics[width=0.99 \columnwidth]{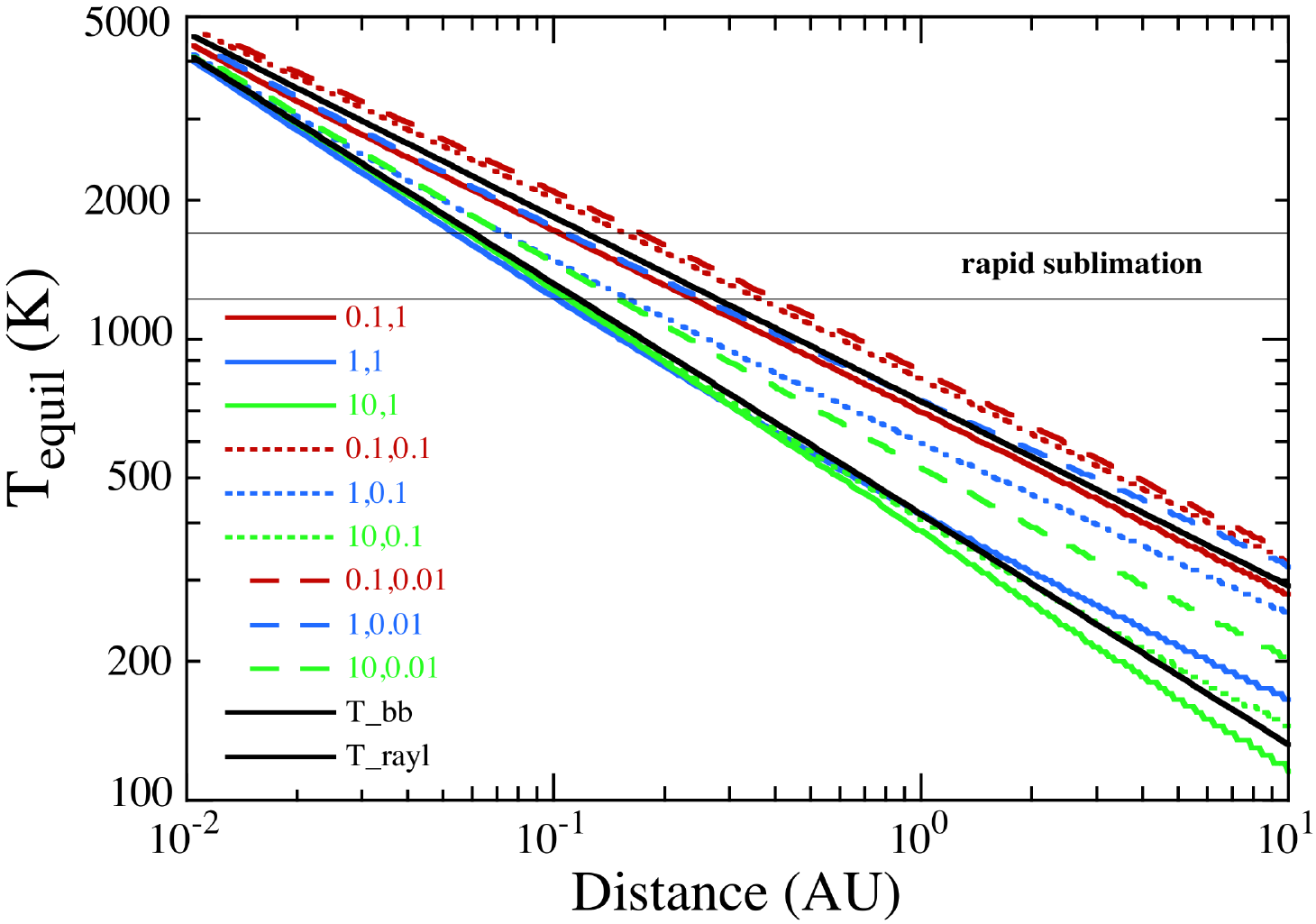}
   \caption{Dust-grain equilibrium temperatures vs.~distance from the host star KIC 3542116 for three different size particles, $s=0.1,1$ and 10 $\mu$m, and three different imaginary parts of their indices of refraction, $k = 1$, 0.1 and 0.01, as indicated in the legend (the first of the two numbers is $s$ and the second is $k$).  The bottom black curve is the idealized equilibrium temperature if the particles absorb and emit as blackbodies, while the upper black curve is ``$T_{\rm rayl}$", as defined in the text.}
   \label{fig:tequil}
\end{figure}

The results for $T_{\rm equil}$ are shown in Fig.~\ref{fig:tequil} for distances ranging from 10 AU in to 0.01 AU from KIC 3542116.  We also show a curve for $T_{\rm equil,bb}$ of the dust particles if they emitted and absorbed as blackbodies:
\begin{equation}
T_{\rm equil,bb} \simeq \sqrt{\frac{R}{2d}} \, T_{\rm eff} \simeq 420 \, \left(\frac{d}{\rm AU}\right)^{-1/2}  K
\end{equation}
For small scaled particle sizes (and/or long wavelengths), $X \equiv 2 \pi s/\lambda$, where $\lambda$ is the wavelength of the light that is interacting with the grain, the normalised absorption cross sections scale simply as $X$. In that case, it has been shown (see, e.g., Rappaport et al.~2014a) that 
\begin{equation}
T_{\rm equil,rayl} \simeq \left(\frac{R}{2d}\right)^{-2/5} T_{\rm eff} \simeq 730 \, \left(\frac{d}{\rm AU}\right)^{-2/5}  K
\end{equation}
Here we retain the nomenclature of the ``Rayleigh" temperature following Xu et al.~(2017).  This higher $T_{\rm equil}$ simply reflects the fact that the particles are better absorbers of the starlight than emitters in the IR, at the same particle size, and therefore the equilibrium temperatures that are attained are higher.  As can be seen from Fig.~\ref{fig:tequil}, $T_{\rm equil,bb}$ adheres closely to the calculated values of $T_{\rm equil}$ for the larger particles, while the $T_{\rm equil,rayl}$ values are higher and closer to the calculated values of $T_{\rm equil}$ for the smaller particles,
 
The bottom line of these calculations is that for many common minerals (e.g., obsidian, magnetite, SiO, fayalite, enstatite, forsterite, corundum, and SiC), which commence rapid sublimation at $T_{\rm equil} \gtrsim 1200-1700$ K\footnote{An exception is magnetite which sublimates at a substantially lower temperature.}, we can estimate that the dust tails begin to sublimate away at distances from KIC 3542116 of $\lesssim 0.1-0.3$ AU (see also Beust et al.~1998).

\section{Another Potential Exocomet Candidate KIC 11084727}
\label{sec:KIC11084727}

Of all the {\em Kepler} target stars that were visually examined, the most compelling case for exocomet transits was KIC 3542116, discussed extensively above.  However, there was one other target star, KIC 11084727, which exhibited a single transit event that was very similar to the three deeper transits found in KIC 3542116.

\begin{figure}
   \centering
   \includegraphics[width=0.99 \columnwidth]{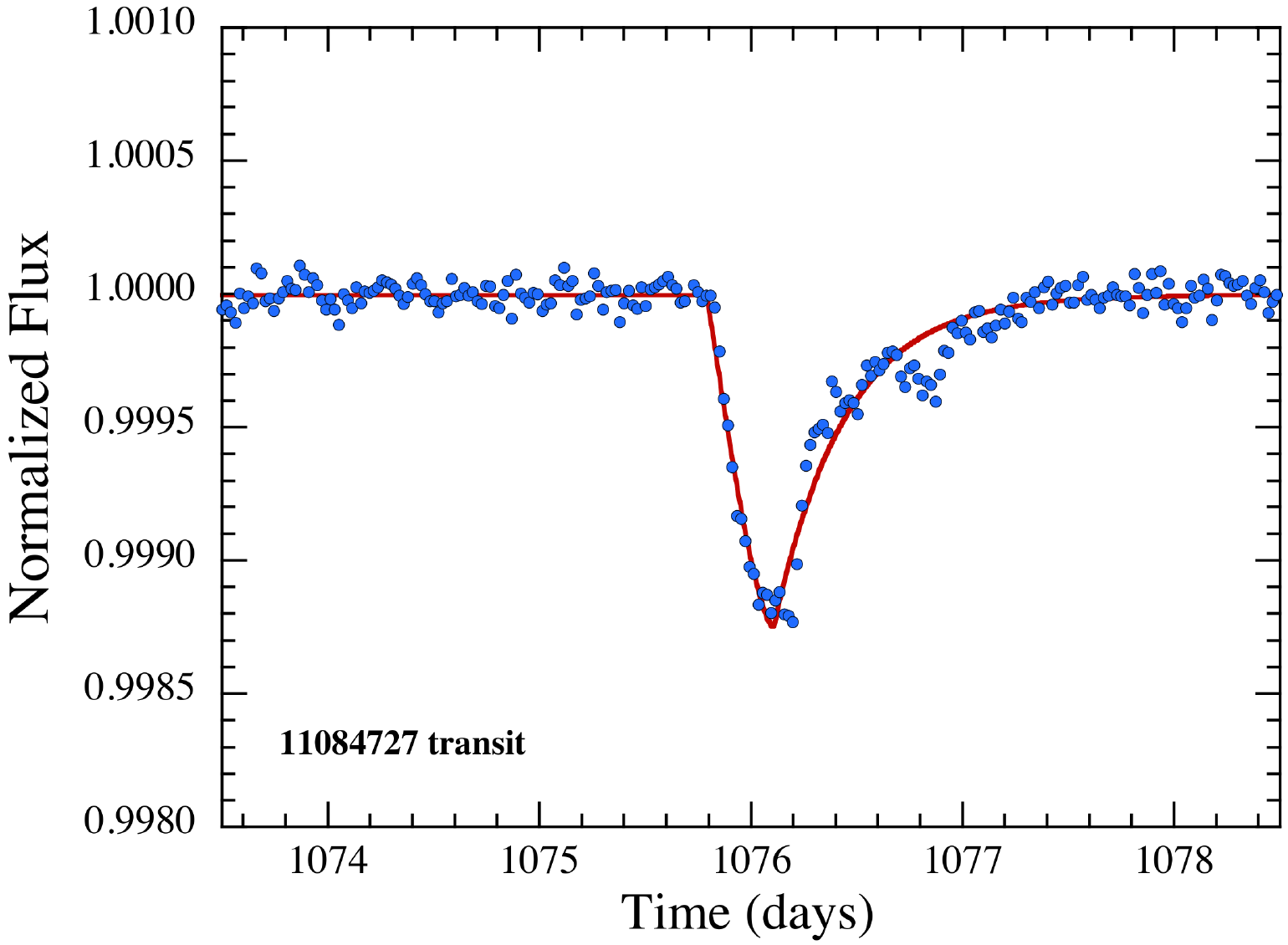}
   \caption{Exocomet-like dip feature in KIC 11084727.  The solid red curve is a model fit similar to those described in Sect.~\ref{sec:model} and shown in Fig.~\ref{fig:transit_profiles}. The shape, depth, and duration of this transit is quite similar to those of transits D992, D1176, and D1268 seen in KIC 3542116.}
   \label{fig:11084727}
\end{figure}

The transit event in KIC 11084727 is shown in Fig.~\ref{fig:11084727} along with a model fit. As is evident from a casual visual inspection, and from the formal model fit (the red curve in Fig.~\ref{fig:11084727}), the transit properties are very similar in shape, depth, and duration to those listed in Table \ref{tbl:parms} for the deeper dips in KIC 3542116. 

The data validation process for this target was essentially the same as described in Sect.~\ref{sec:validation} for KIC 3542116.  All indications are that this dip is astrophysical in origin and is associated with KIC 11084727.

The target star KIC 11084727 is a near twin to KIC 3542116 as can be seen from a compilation of their photometric properties in Table \ref{tbl:mags}, with nearly identical magnitudes (at Kp = 9.99), similar $T_{\rm eff}$ values, and comparable radii. The similarities between \thisstar\ and KIC 11084727 are particularly striking given that the majority of \Kepler\ targets were cooler, Sun-like stars and suggest that comet transits may preferentially happen around stars of this spectral type. 

The fact that we have detected two individual stars with similar comet-like transit events also suggests that there may be more (perhaps shallower) comet-like transits hidden in the \Kepler\ dataset. 

\section{Discussion}
\label{sec:discuss}

In this section we discuss some possible follow-up observations of KIC 3542116 and KIC 11084727 that may connect these systems to other exocomet systems found with ground-based observations.  We also attempt to understand the relative detection sensitivity of dusty transits using photometry and spectral line changes.  A number of dynamical effects that might be responsible for driving comets into orbits close to the host star are briefly discussed.  Finally, we compare our two systems with Boyajian's star (KIC 8462852). 

It might be profitable to carry out follow-up ground-based spectroscopic studies of KIC 3542116 to see if any of the same type of spectral line changes such as are found in $\beta$ Pic, 49 Ceti, HD 42111, HD 172555, and $\phi$ Leo can be discerned in KIC 3542116. Of the 16 stars listed by Welsh \& Montgomery (2015) as exhibiting spectral line changes that are likely due to exocomets, the magnitudes range from 3.6 to 7.2 with a mean of 5.6. Moreover, the spectral types of these stars range from B9 to F6, with 2 of the 23 being stars of the F spectral type. Thus, aside from the fact that KIC 3542216 and KIC 11084727 are more than an order of magnitude fainter than these other stars, it may still be possible, even if challenging, to monitor the spectral line shapes of KIC 3542116 and KIC 11084727 for changes. We believe that connecting photometric transits to spectral transits in the same star could prove very rewarding.
    
Neither KIC 3542116 nor KIC 11084727 shows any particular evidence for being extremely young, e.g., via very rapid rotation or WISE excess flux.  Nor is there  any specific reason to believe that there is disk activity or populations of minor bodies at orbital separations much greater than these inferred for the comets in this work.  Such debris might be expected to exhibit CO emission as is seen in HD 181327 (Marino et al.~2016), Eta Corvi (Marino et al.~2017), and Fomalhaut (Matr\`a et al.~2017).  Nevertheless, the stars reported on herein, are sufficiently unusual among the {\em Kepler} ensemble of $2 \times 10^5$ stars, that it could be worth the gamble to use ALMA to search for CO emission around KIC 3542116 and KIC 11084727. 

The observations of likely comets in two {\em Kepler} stars in this work raise some interesting questions by way of comparison with the comets (FEBs) inferred from spectral-line changes in a substantial number of primarily bright A stars (e.g., Beust et al.~1990; Welsh \& Montgomery 2015). In particular, why are we not detecting `swarms' of comets as are suggested by the papers on FEBs? 

Of the $2 \times 10^5$ {\em Kepler} stars studied continuously for approximately four years, we have found only 6 comet-like dips in one star (KIC 3542116) and 1 dip in KIC 11084727.  Presumably, {\em Kepler} is not as sensitive to small comets as the spectroscopic methods are. In the spectral approach, the comet is only blocking a small part of the light (a fraction of one spectral line), but which can be readily detectable in the line profile. By contrast, when looking at transits in {\em Kepler} data, we are collecting much of the bolometric flux from the star. Therefore, a much larger comet may be required to be detectable by {\em Kepler} than by a spectrograph on a large telescope. 

More quantitatively, we can write the {\em total} mass loss rate of a comet crossing the stellar disk of the host star as
\begin{equation}
\dot M \simeq \Sigma_j  w_j v_{j \perp} f_j^{-1}
\label{eqn:mdot}
\end{equation}
where $\Sigma_j$ is the observed mass column density of component, $j$ (e.g., dust, CaII, MgII, FeII), assumed uniform over a strip of width $w_j$ that extends across the disk of the stellar host; $v_{j \perp}$ is the material outflow velocity, i.e., relative to the comet, and perpendicular to the line of sight; and $f_j$ is the mass fraction of component $j$ in the comet effluents.  We can write Eqn.~(\ref{eqn:mdot}) more specifically as applied to the dust case (in regard to the {\em Kepler} dips):   
\begin{eqnarray}
\dot M & \simeq & 2\rho_d s_d D_d  R_* v_\perp  f_d^{-1} \nonumber \\
& \simeq & 3 \times 10^{11} \left[\frac{D_d}{0.001} \right] \left[\frac{v_\perp}{5\,{\rm km/s}}\right] \left[\frac{f_d}{0.1}\right]^{-1} {\rm g~s}^{-1}
\label{eqn:mdot1}
\end{eqnarray}
and for the case of CaII ions as observed in the FEB sources:
\begin{eqnarray}
\dot M & \simeq & 2 \mu_{\rm CaII}  N_{\rm CaII} R_* v_\perp  f_{\rm CaII}^{-1} \nonumber \\
& \simeq & 10^{11} \left[\frac{N_{\rm CaII}}{10^{12}/{\rm cm}^{2}}\right] \left[\frac{v_\perp}{50\,{\rm km/s}}\right]\left[\frac{f_{\rm CaII}}{0.001}\right]^{-1} \! \!\!{\rm g~s}^{-1}
\label{eqn:mdot2}
\end{eqnarray}
where the subscripts ``$d$'' and ``CaII'' refer to dust and the CaII ion, respectively.  The other symbols not yet defined are: $s_d$ the characteristic size of the dust grains (taken to be 1 $\mu$m); $D_d$ the minimum depth of a comet dip that can be detected with {\em Kepler}; $\mu_{\rm CaII}$ the atomic weight of a Ca atom; and $N_{\rm CaII}$ is the {\em average} column number density of CaII as seen by a distant observer.  In the second line of both equations, the values of some of the important parameters are normalised to suggestive illustrative values.  The limiting column density was taken from Hobbs et al.~(1985) for the case of the bright star $\beta$ Pic at $V \simeq 3.8$; for stars that are up to a couple of magnitudes fainter, we assume that the sensitivity limit is only a few times higher, which would make the leading coefficient comparable to that for dust transits.  

The conclusion we draw from these expressions is that ground-based spectral observations of bright stars (magnitude 2-6) should be more sensitive, in terms of detecting a given $\dot M$, than are the {\em Kepler} observations, but with substantial uncertainties in the choice of parameter values.  One caveat is in order, however, when interpreting equations (\ref{eqn:mdot1}) and (\ref{eqn:mdot2}). Presumably the dust will only survive rapid sublimation at distances beyond $\sim$0.1-0.3 AU (see Fig.~\ref{fig:tequil}).  By contrast, the atoms (in particular CaII, MgII, and FeII) will mostly be present closer in where the dust, bearing many of the heavier elements, sublimates and the minerals become photo-dissociated and ionised. 

With regard to the number of comets that should be seen crossing the disk of the host star, this rate should depend heavily on whether the infalling comet orbits are distributed roughly isotropically (lower rate) or if the reservoir of bodies producing the dusty tails has orbits that are coplanar with the angular momentum vector of the system (higher rate if the observer lies in this plane).  We do have some limited information on the viewing inclination angle with respect to the spin axis of KIC 3542116.  From Table \ref{tbl:mags} we find $v \, \sin \, i \simeq 57.3 \pm 0.3$ km s$^{-1}$, $R_{\rm *} \simeq 1.56 \pm 0.15 \, R_\odot$, and a rotation frequency of $0.888 \pm 0.04$ cycle d$^{-1}$.  When we use this information to compute the inclination angle, $i$, we find that $45^\circ \lesssim i \lesssim 80^\circ$ with 95\% confidence.  This is suggestive that we could be viewing the system from at least a partially favorable in-plane vantage point.  It will be helpful to firm up these uncertainties in future work.

In Sect.~\ref{sec:orbits} we made some initial assessments of the kinds of orbits that were most likely responsible for the exocomet transits we report (see, in particular, the right panel of Fig. 11). There are basically two major dynamical mechanisms for generating potentially large numbers of transiting exocoments.   These have been very well explored in the context of the best-studied FEB system -- $\beta$ Pictoris.  However, we should keep in mind that in $\beta$ Pic, there is a high preponderance of red-shifted FEB events, which implies a particular orientation for the comet trajectories.  With this caveat in mind, we note that these dynamical mechanisms involve secular perturbations by a distant planet. First, they may be generated via the Kozai-Lidov mechanism (see, e.g., Bailey et al.~1992).  The second mechanism involves resonances, either secular (Levison et al.~1994), or mean-motion (Beust \& Morbidelli 1996; 2000). In the former case the exocomet orbits should be roughly isotropically distributed thanks to a rotational invariance of the Kozai Hamiltonian.  Conversely, in the latter case the longitude of periastron of the perturbing planet controls the geometry of the infall.  Also, in this case Beust \& Valiron (2007) showed that the exocomets may have large inclination oscillations when reaching the FEB state even if they started out with only modest inclinations with respect to the orbit of the perturbing planet.  We note that in the solar system, most sun-grazers are thought to arise from the Kozai mechanism (e.g., Bailey et al.~1992).  In the case of $\beta$ Pic, the mean-motion-resonance mechanism was favoured to match the abundant statistics of the FEB events in that system.  By contrast, for KIC 3542116, with only a few transit events detected, all of the above mechanisms are worthy of consideration.

Finally, the deep dips in the flux of KIC 8462852 (aka `Boyajian's Star'; Boyajian et al.~2016) are worth trying to relate to what is observed in KIC 3542116.  By contrast, the largest flux dips in the former star reach 22\% which is more than two orders of magnitude greater than the transits we see in KIC 3542116.  Furthermore, the dips in KIC 8462852 can last for between 5 and 50 days, depending on how the beginning and end points of the dip are defined.  These are one to two orders of magnitude longer than for the transits in KIC 3542116. Finally, we note that none of the dips in KIC 8462852 has a particularly comet-shaped profile.  There have been a number of speculations about the origin of the dips in KIC 8462852, including material resulting from collisions of large bodies and moving in quasi-regular orbits (Boyajian et al.~2016); swarms of very large comets (Boyajian et al.~2016); and even a ring of dusty debris in the outer solar system (Katz 2017). However, there is currently no compelling evidence for any of these scenarios.

\section{Summary and Conclusions}
\label{sec:summary}

In this work we reported the discovery of six apparent transits in KIC 3542116 that have the appearance of a trailing dust tail crossing the disk of the host star. We have tentatively postulated that these are due to between 2 and 6 distinct comet-like bodies in the system.  We also found a single similarly shaped transit in KIC 11084727. Both of these host stars are of F2V spectral types.  

We have carefully vetted the data from these target stars, including assessing the difference images in and out of transit, analyzing potential video crosstalk, and inspecting the data quality flags associated with the dip events.  The vetting also included deep high-resolution imaging studies of our prime target, KIC 3542116.  No companion stars were found within an outer working angle of $\rho \sim 5$\arcsec, though the nearby star KIC 3542117 ($\rho = 10$\arcsec) might plausibly be a bound companion. 

The spot rotation period in KIC 3542116 is about as long as the durations of the deeper transits we see, and therefore it is difficult to imagine that they are caused by highly variable spots (which tend to produce dips at a fraction of the rotation period).  Nonetheless, we cannot categorically rule out the possibility that the transit-like events are caused by some previously unknown type of stellar variability.  

With this caveat in mind, we proceeded to study these systems under the assumption that the dips in flux are indeed due to dusty-tailed objects transiting the host stars.
We then fit these transits with a model dust tail that is assumed to have an exponentially decaying extinction profile.  The model profiles fit the transits remarkably well.

The inferred speeds of the underlying dust-emitting body during the transits are in the range of $35-50$ km s$^{-1}$ for the deeper transits in KIC 3542116 and for the single transit in KIC 11084727. For the more shallow and narrow transits in KIC 3542116, the inferred speeds are $75-90$ km s$^{-1}$. From these speeds we can surmise that the corresponding orbital periods are $\gtrsim 90$ days (and most probably, much longer) for the deeper transits, and $\gtrsim 50$  days for the shorter events.
    
Solar system sun-grazing comets typically have extremely long orbital periods (e.g., $\sim$2300 years for the members of the populous Kreutz group).  Halley's comet, which has an apohelion distance of $\sim$125 $R_\odot$, has a period of $\simeq 75$ years, while the shortest known period for a bright comet is Comet Encke at 3.3 years.  The overall shortest period is Comet 311/Pan-STARRS with a period of 3.2 years.  Thus, if either the three deeper or the three more shallow transits in KIC 3542116 are from single orbiting bodies, then the maximum associated periods of 51-92 days would be considerably shorter than for Solar-system comets.  The periods of the comets producing the FEB events are largely unknown.  However, the characteristic infall velocities associated with the CaII FEB's ($\sim$50 km s$^{-1}$; e.g., Beust et al.~1990) are compatible with what we find for KIC 3542116 and KIC 11084727.
    
The fact that we find comet-like transits in two {\em Kepler} target stars holds out the promise that such events are not particularly rare. This is especially true when we note that the survey was made visually and without the aid of a computer search. In turn, the fact that the search was carried out visually raises the issue of its completeness. In this regard, we believe that there was no particular obstacle to finding asymmetric transits with depths of $\gtrsim$ 0.1\% and lasting for $\gtrsim$ 1/3 day, even in the presence of significant star-spot activity. Furthermore, we likewise found that data breaks and artefacts would also not have impeded the search. 
    
We thus believe that we have found the majority of such comet-like transits in the {\em Kepler} data set, though we cannot preclude the possibility that there are many more such features with depths $\lesssim$ 0.1\%.

We reiterate that there are striking similarities between KIC 3542116 and KIC 11084727 in terms of both the stellar properties and the comet-like transit events. This is also noteworthy because the majority of {\em Kepler} targets were cooler, Sun-like stars.  This might suggest that comet transits may preferentially happen around stars of this spectral type, and it would be instructive to try to understand why this might be.

One encouraging note in regard to finding more such comet-like transits in other stars is that dips in flux at the $\gtrsim$ 0.1\% level and lasting for hours to days should not be particularly challenging for the photometry in the upcoming TESS mission (Ricker et al.~2015). Furthermore, the host stars are likely to be bright, plausibly even brighter than the 10th magnitude stars reported on here.

{\bf Note added in manuscript}: After this manuscript was submitted we became aware of a remarkably prescient paper by Lecavelier des Etangs, Vidal-Madjar, \& Ferlet (1999) which predicted the photometric profiles of exocomet dust-tail transits of their host star (see also Lecavelier des Etangs 1999).  The calculated profiles look rather remarkably like the ones we have found and reported on here.  Therefore, it appears that this current work could help to confirm these earlier predictions, and similarly the predictions may help strengthen the case that we have indeed observed exocomet transits.

\acknowledgments
We are especially grateful to the referee, Herv\'e Beust, for a substantial number of very instructive suggestions leading to the improvement of the manuscript.  We thank Tabetha Boyajian, Bradley Schaefer, and Benjamin Montet for discussions of the long-term photometric variations in KIC 3542116.  We also appreciate helpful discussions of comet properties with Nalin Samarasinha and Bruce Gary. A.\,V.~was supported by the NSF Graduate Research Fellowship, grant No.~DGE 1144152, and also acknowledges partial support from NASA's TESS mission under a subaward from the Massachusetts Institute of Technology to the Smithsonian Astrophysical Observatory, Sponsor Contract Number 5710003554.  This work was performed in part under contract with the California Institute of Technology (Caltech)/Jet Propulsion Laboratory (JPL) funded by NASA through the Sagan Fellowship Program executed by the NASA Exoplanet Science Institute.  This research has made use of NASA's Astrophysics Data System and the NASA Exoplanet Archive, which is operated by the California Institute of Technology, under contract with the National Aeronautics and Space Administration under the Exoplanet Exploration Program. This paper includes data collected by the \Kepler\ mission. Funding for the \Kepler\ mission is provided by the NASA Science Mission directorate. Some of the data presented in this paper were obtained from the Mikulski Archive for Space Telescopes (MAST). STScI is operated by the Association of Universities for Research in Astronomy, Inc., under NASA contract NAS5--26555. Support for MAST for non--HST data is provided by the NASA Office of Space Science via grant NNX13AC07G and by other grants and contracts.  

Facilities: \facility{Kepler/K2, FLWO:1.5\,m (TRES), Keck Observatory}


\clearpage

\end{document}